\documentclass[11pt]{article}
\pdfoutput=1
\usepackage{amsmath,amssymb,mathtools}
\usepackage{graphicx}
\usepackage{units}
\usepackage{url}
\usepackage{booktabs}
\usepackage{subfigure}
\usepackage{hyperref}
\usepackage{empheq}

\usepackage[sc,small]{caption}
\usepackage{color}

\DeclareFontFamily{U}{mathx}{\hyphenchar\font45}
\DeclareFontShape{U}{mathx}{m}{n}{<-> mathx10}{}
\DeclareSymbolFont{mathx}{U}{mathx}{m}{n}
\DeclareMathAccent{\wb}{0}{mathx}{"73}

\newcommand{\mathscr}{\mathcal}

\newcommand{\be}{\begin{eqnarray}}
\newcommand{\ee}{\end{eqnarray}}
\newcommand{\bea}{\begin{eqnarray}}
\newcommand{\eea}{\end{eqnarray}}

\newcommand{\beqn}{\begin{eqnarray}}
\newcommand{\eeqn}{\end{eqnarray}}

\begin{document}

\hfill LMU-ASC 19/18\\
\begin{center}
{\bf\LARGE
Field redefinitions and K\"ahler potential in string theory at 1-loop}

\vspace{1.5cm}
{\large
{\bf Michael Haack$^{\dag}$, }
{\bf Jin U Kang$^{\star}$}
\vspace{1cm}

{\it
$^{\dag}$ 
 Arnold Sommerfeld Center for Theoretical Physics \\ 
Ludwig-Maximilians-Universit\"at M\"unchen \\ 
Theresienstrasse 37, 80333 M\"unchen, Germany\\ [5mm] 
$^{\star}$ 
Abdus Salam International Centre for Theoretical Physics\\  Strada Costiera 11, Trieste 34014, Italy 
\\and\\
Department of Physics, Kim Il Sung University\\ RyongNam Dong, TaeSong District, Pyongyang, DPR.\ Korea 
 \\[5mm]
}
}
\end{center}
\vspace{2mm}

\begin{center}
\end{center}

Field redefinitions at string 1-loop order are often required by supersymmetry, for instance in order to make the K\"ahler structure of the scalar kinetic terms manifest. We derive the general structure of the field redefinitions and the K\"ahler potential at string 1-loop order in a certain class of string theory models (4-dimensional toroidal type IIB orientifolds with ${\cal N}=1$ supersymmetry) and for a certain subsector of fields (untwisted K\"ahler moduli and the 4-dimensional dilaton). To do so we make use of supersymmetry, perturbative axionic shift symmetries and a particular ansatz for the form of the 1-loop corrections to the metric on  the moduli space. Our results also show which terms in the low-energy effective action have to be calculated via concrete string amplitudes in order to fix the values of the coefficients (in the field redefinitions and the K\"ahler potential) that are left undetermined by our general analysis based on (super)symmetry. 

\clearpage
 
\tableofcontents

\clearpage


\section{Introduction}

The low-energy effective field theory of explicit string theory models is an important ingredient in attempts to make contact between string theory and phenomenology. In this paper we consider a subsector of the low-energy effective theory of 4-dimensional toroidal type IIB orientifolds with minimal supersymmetry, consisting of the kinetic terms of the diagonal untwisted K\"ahler moduli (i.e.\ the moduli determining the volumes of the three $\mathbb{T}^2$-factors of the internal space) and the 4-dimensional dilaton.\footnote{For many toroidal orientifolds, only the diagonal K\"ahler moduli survive in the untwisted sector. However, especially for $\mathbb{Z}_3$- and $\mathbb{Z}_4$-orientifolds and also some $\mathbb{Z}_6$-orientifolds, one has $h^{1,1}_{\rm untw.}>3$, cf.\ table 20 in \cite{Blumenhagen:2006ci}.} These fields play an important role in many approaches to model building and the goal of the present work is to gain a better understanding of this sector at string 1-loop order. In particular we focus on two questions:
\begin{itemize}
\item What is the K\"ahler potential of the untwisted K\"ahler moduli and the 4-dimensional dilaton at 1-loop, consistent with ${\cal N}=1$ supersymmetry and shift symmetries?
\item What are the 1-loop field redefinitions of these fields?
\end{itemize}
The necessity for a field redefinition at 1-loop order arises because in general the metric on the sigma model target space ceases to be manifestly K\"ahler, i.e.\ it ceases to be given by the second derivative of a real function with respect to the original (complex) field variables, once 1-loop corrections to the metric are taken into account.\footnote{We wonder if this could be dealt with by redefining the vertex operators at loop-level, as in section 7.6.3 in \cite{Witten:2012bh}.} Consequently, it is also impossible to read off the structure of the K\"ahler potential without first finding a field basis for which the K\"ahlerness of the metric is manifest. Thus, the two points above are intimately related and have to be solved simultaneously. 

The need for field redefinitions in the context of (toroidal) type IIB orientifolds has also been observed at disk level (i.e.\ at order $e^{\Phi}$ relative to the leading form of the field definitions, where $\Phi$ is the dilaton). This is due to the presence of the open string sector and can either be inferred by an analysis of the kinetic terms resulting from a Kaluza-Klein reduction of the coupled supergravity and DBI actions \cite{Antoniadis:1996vw,Camara:2003ku,Grana:2003ek,Jockers:2004yj,Grimm:2008dq,BerasaluceGonzalez:2012vb}, by analyzing the physical gauge couplings \cite{Antoniadis:1999ge,Berg:2004ek,Conlon:2009xf,Conlon:2009kt} or by considering the transformation of the field variables under discrete shifts of the open string fields \cite{Camara:2009uv,Carta:2016ynn}. Almost all the field redefinitions of the untwisted K\"ahler moduli and the dilaton observed in these papers vanish, however, when the open string scalars are set to zero.\footnote{The field redefinition discussed in \cite{Conlon:2009xf} is an exception.} Field redefinitions can also already arise at sphere level and disk level from $\alpha'$-corrections. Examples of this were discussed in \cite{Antoniadis:2003sw,Grimm:2013gma,Grimm:2013bha}. We will, however, restrict our analysis to field redefinitions arising at string 1-loop order.

Field redefinitions at string 1-loop order (i.e.\ at order $e^{2 \Phi}$ relative to the leading form of the field definitions) were discussed much less in the literature. A well-known example arises for the dilaton in the heterotic string which was first discussed in \cite{Derendinger:1991hq}. Examples in the context of type I and type II models (even though with ${\cal N}=2$ supersymmetry) were discussed in \cite{Antoniadis:1996vw,Antoniadis:2003sw,Berg:2005ja}. To our knowledge, 1-loop field redefinitions in type II orientifolds with ${\cal N}=1$ supersymmetry have not been studied so far. 

The importance of the first point of the above list for string model building should be rather obvious. String loop corrections to the K\"ahler potential were discussed in the context of moduli stabilization (see \cite{vonGersdorff:2005bf,Berg:2005yu,Cicoli:2008va,Antoniadis:2018hqy} for examples in type IIB compactifications) and in approaches to inflation within string theory, cf.\ \cite{Baumann:2014nda} for an overview. For instance, loop corrections play an important role in fibre inflation, introduced in \cite{Cicoli:2008gp}. However, also the second point might have interesting phenomenological consequences. Redefinitions of K\"ahler moduli by open strings were instrumental in attempts to embed inflation into string theory, cf.\ \cite{Kachru:2003sx}, and the redefinition of the volume moduli at 1-loop level (even though including blow-up modes) could have some noticeable effect on the phenomenology of the Large Volume scenario, cf.\ \cite{Conlon:2010ji}.

Our strategy is to derive and solve general constraints arising from  ${\cal N}=1$ supersymmetry (i.e.\ from the fact that the moduli metric is K\"ahler), from axionic shift symmetries of the moduli metric and, finally, from a well motivated ansatz for the moduli metric (cf.\ \eqref{Gtt_N=1}, \eqref{Gcc_N=1}, \eqref{Gtt_N=2} and \eqref{Gcc_N=2} below). This allows us to determine the general structure of the K\"ahler potential and the field redefinitions of the untwisted K\"ahler moduli and the dilaton, compatible with the above three constraints. This analysis does not allow us to fix certain coefficients whose determination requires explicit string calculations (which we leave for future work). In spirit, our analysis bears some similarity to the strategy followed in \cite{Antoniadis:2003sw}. There the authors also determined general constraints on the form of the metric of the universal hypermultiplet in type II compactifications, arising from ${\cal N}=2$ supersymmetry and shift symmetries. To fix the final form of the metric a string calculation was necessary. Also in their case, supersymmetry required a redefinition of the volume modulus at 1-loop order, where it mixes with the 4-dimensional dilaton. 

The determination of the 1-loop K\"ahler potential for the untwisted K\"ahler moduli and the dilaton in certain ${\cal N}=1$ type IIB orientifolds (a $\mathbb{Z}_2 \times \mathbb{Z}_2$- and a $\mathbb{Z}_6'$-orientifold) was already undertaken in \cite{Berg:2005ja,Berg:2014ama} and our findings concerning the K\"ahler potential are consistent with the earlier results. However, our results are more general, as they are valid for an arbitrary 4-dimensional toroidal type IIB orientifold with minimal supersymmetry.\footnote{We are always using the language of type IIB orientifolds with D9/D5-branes. However, our final results for the 1-loop field redefinitions and the correction to the K\"ahler potential should also be valid for type IIB orientifolds with D3/D7-branes, with the appropriate definitions of the tree level moduli fields. This will be discussed further at the end of section \ref{effectiveaction}.} Moreover, our findings for the field redefinitions are new. Like the present paper, also \cite{Berg:2005ja} determined the general structure of the K\"ahler potential for the models under scrutiny (except for the ${\cal N}=2$ case of type I compactified on $\mathbb{T}^2 \times {\rm K}3$, where the additional supersymmetry fixed also the coefficient of the 1-loop correction to the K\"ahler potential). The analysis of the $\mathbb{Z}_6'$-orientifold was based on T-duality arguments which did not fix certain coefficients in the K\"ahler potential and, in particular, it did not give any hint towards the field redefinitions required at 1-loop level. The field redefinitions, on the other hand, are important ingredients in the final determination of the 1-loop K\"ahler potential as they can lead to the absorption or additional generation of 1-loop contributions to the moduli kinetic terms. Hence, we consider our results concerning the K\"ahler potential as a nice check of the consistency of \cite{Berg:2005ja}.

Thus, even though partial results in certain individual models were available in the literature, our findings allow for a more rigorous and general understanding of the 1-loop structure of the moduli K\"ahler potential (in the mentioned subsector of the fields). Our results show how the fields should be redefined in order for the different terms in the moduli metric to be consistent with ${\cal N}=1$ supersymmetry and axionic shift symmetries. Moreover, our results indicate clearly which quantities in the low-energy effective action one would have to calculate via string amplitudes in order to fix the undetermined coefficients in the K\"ahler potential and the field redefinitions. This paves the way for a complete determination of the K\"ahler potential at 1-loop by concrete string amplitude calculations. Somewhat surprisingly, we found that only very few quantities have to be calculated by string theory in order to determine the 1-loop correction to the K\"ahler potential (for instance, for $\mathbb{Z}_N$ orientifolds a single component of the moduli metric in Einstein frame is sufficient). This is due to the fact that the (super)symmetries of the low-energy effective action lead to relations between different components of the moduli metric which considerably simplify the task of determining the 1-loop correction to the K\"ahler potential. 

The organization of the paper is as follows. In section \ref{effectiveaction} we begin with a review of some relevant aspects of the low-energy effective action of toroidal type IIB orientifolds, focusing on the kinetic terms and on the definition of the field variables for which the metric on moduli space is K\"ahler at tree-level (always in the subsector of fields that we are considering, cf.\ the beginning of this introduction). In section \ref{redef} we then discuss the general framework for obtaining the field redefinitions and the form of the K\"ahler potential at 1-loop order, imposing supersymmetry and axionic shift symmetries. We use this framework in sections \ref{N=1} and \ref{sec.N=2} in order to obtain the general structure of the contributions to the K\"ahler potential and the field redefinitions arising from the ${\cal N}=1$ and ${\cal N}=2$ sectors of an arbitrary 4-dimensional toroidal type IIB orientifold with minimal supersymmetry.  Section \ref{observation} contains an observation on the structure of the field redefinitions and the corrections to the K\"ahler potential. We apply the results of section \ref{concretecomps} to the example of the $\mathbb{Z}_6'$-orientifold in section \ref{applicationZ6}. Finally, we end with concluding remarks in section \ref{summary}. The appendix contains some technical aspects of our calculations and also an application of our methods to the ${\cal N}=2$ theory of type I compactified on $\mathbb{T}^2 \times {\rm K}3$. 

We framed some key equations and results for an easier orientation of the reader.


\section{String 1-loop effective action}
\label{effectiveaction}

We consider the kinetic terms of the 4-dimensional dilaton and the volume moduli of the three 2-tori in an arbitrary 4-dimensional and minimally supersymmetric toroidal type IIB orientifold, together with their axionic partners. At tree-level and in Einstein-frame they look like\footnote{In this paper superscripts $(0)$ and $(1)$ denote tree-level quantities and string 1-loop corrections, respectively. The reader might wonder why we include a superscript $(0)$ only for the $c$-variables and not for the $t$-variables in \eqref{tree-EF-action-string}. The reason is the following: In section \ref{redef} we are going to discuss possible redefinitions of the K\"ahler variables (i.e.\ the variables for which the scalar metric is given by the second derivative of a K\"ahler potential). This might become necessary at 1-loop order if the corrected metric can not be written anymore as the second derivative of a corrected K\"ahler potential with respect to the tree-level variables. However, whereas the $c^{(0)}$-variables are in fact the real parts of the K\"ahler variables at tree-level, the $t$-variables are not their imaginary parts. These are rather given by the $\tau^{(0)}$-variables introduced below in \eqref{taut}. Hence it is the $\tau$-variables that are going to be redefined at 1-loop order and not the $t$-variables and, thus, we do not have to indicate their tree-level form with a superscript $(0)$.} 
\be \label{tree-EF-action-string}
S_4 &=&\frac{1}{\kappa_4^2}  \int d^4 x \sqrt{-g} \left[  {1 \over 2 } R  
 - \sum_{i,j=0}^3  \left( G^{(0)}_{t_i t_j} \, \partial_\mu t_i \partial^\mu t_j
+ G^{(0)}_{c_i^{(0)} c_j^{(0)}}\,  \partial_\mu c_i^{(0)} \partial^\mu c_j^{(0)} \right) \right] + \ldots \ .
\ee
Here, $\kappa_4^2$ is the 4-dimensional gravitational constant, $t_0$ denotes the 4-dimensional dilaton, 
\be
t_0 = \Phi_4\ ,
\ee
and the $t_i$ $(i \in \{ 1,2,3 \})$ are the volumes of the 2-tori of the toroidal ${\cal N} = 1$-orientifold, measured with the string frame metric. The tree-level metric for these fields is \cite{Antoniadis:1996vw,Aldazabal:1998mr}
\be \label{treelevelmetrics}
G^{(0)}_{t_0 t_0}=1\ , \quad G^{(0)}_{t_i t_j} = \frac{1}{4 t_i^2} \delta_{ij}\ , \quad i,j \in \{ 1,2,3 \} \ . 
\ee
As mentioned in the introduction, we focus on field-redefinitions induced by string 1-loop effects and, thus, for most part of the paper we do not consider any $\alpha'$-corrections to the tree-level metric, as the one of \cite{Becker:2002nn}. Including those would lead to additional terms in the 1-loop field redefinitions which are doubly suppressed, in $g_s$ and the inverse overall volume ${\cal V}^{-1}$.\footnote{In order to avoid misunderstandings, let us reiterate that $\alpha'$-corrections can not only contribute to the 1-loop field redefinitions but can also induce field redefinitions already at tree level, as in \cite{Antoniadis:2003sw,Grimm:2013gma,Grimm:2013bha}. These tree level field redefinitions would not be doubly suppressed. However, we only consider 1-loop field redefinitions in this paper.} The fields $c^{(0)}$ arise from the RR-sector; $c_0^{(0)}$ is the scalar dual to the 2-form field $C_{\mu \nu}$ (with 4-dimensional indices) and the $c_i^{(0)}$ $(i \in \{ 1,2,3 \})$ are the components of the RR 2-form $C_2$ with indices along the $i$th torus. The metric $G^{(0)}_{c_i^{(0)} c_j^{(0)}}$ is diagonal and will be given below (in \eqref{Gcc0}).

We stress that, at tree-level, it is the form of the action in Einstein-frame given in equation \eqref{tree-EF-action-string} that one obtains by comparing with the string S-matrix elements at sphere level when using the conventional form of the vertex operators for the graviton and dilaton
\be
V(k, \epsilon) = -\frac{2}{\alpha'} \epsilon_{\mu \nu} (i \partial X^\mu + \tfrac12 \alpha' k \cdot \psi\, \psi^\mu)(i \bar \partial X^\nu + \tfrac12 \alpha' k \cdot \tilde \psi\, \tilde \psi^\nu) e^{i k \cdot X(z, \bar z)}\ , \label{vertexhD}
\ee
where the polarisation tensor $\epsilon_{\mu \nu}$ is given by 
\bea
\epsilon^{(h)}_{\mu \nu} & = & \epsilon^{(h)}_{\nu \mu}\ , \quad  \epsilon^{(h)}_{\mu \nu} \eta^{\mu \nu} = 0 = k^\mu  \epsilon^{(h)}_{\mu \nu}\ , \qquad \qquad \qquad  ({\rm graviton}) \label{gravitonpolarisation} \\
\epsilon^{(D)}_{\mu \nu}  & = & \frac{1}{\sqrt{2}} (\eta_{\mu \nu} - k_\mu \bar k_\nu - \bar k_\mu k_\nu)\ , \quad k^\mu  \epsilon^{(D)}_{\mu \nu} = 0 \qquad ({\rm dilaton}) \label{dilatonpolarisation}
\eea
with an auxiliary vector $\bar k _\mu$ that satisfies $\bar k^2 = 0$ and $\bar k \cdot k = 1$, cf.\ eq.\ (16.9) in \cite{Blumenhagen:2013fgp}. This fact is due to the choice of the vertex operators given above, which generate states that are \emph{orthogonal} to each other at tree-level, as explicitly shown e.g.\ in sec. 16.3 of \cite{Blumenhagen:2013fgp}. In \eqref{gravitonpolarisation} $h$ stands for the graviton and in \eqref{dilatonpolarisation} $D$ denotes the fluctuations of the dilaton, 
\be
D \sim \Phi_4 - \overline \Phi_4\ ,
\ee
where, as usual, the constant background value of the dilaton, $\overline{\Phi}_4$, determines the loop counting parameter, $g_s  = e^{\overline \Phi_4}$. 

Now, to the tree-level action \eqref{tree-EF-action-string} we add 1-loop corrections that one could again obtain by matching with string S-matrix elements, i.e.\footnote{We should mention that in writing down \eqref{1-loop-EF-action-string-1} we made an assumption, i.e.\ that the 1-loop 3-point function of two gravitons and a dilaton is not vanishing (leading to the dilaton dependence of the 1-loop correction of the Einstein-Hilbert term). The corresponding amplitude at sphere level vanishes after summing over all kinematical factors and using \eqref{gravitonpolarisation} and \eqref{dilatonpolarisation}, cf.\ section 16.3 in \cite{Blumenhagen:2013fgp}. However, the kinematical factors at 1-loop level and to order ${\cal O}(k^2)$ can be different from the ones at sphere level, given that some of the vertex operators have different picture number. This happens for instance for the 3-point function of gravitons for which the kinematical factor at 1-loop, given in equation (3.2) of \cite{Antoniadis:2002tr}, differs from the one at sphere level, cf.\ equation (16.105) in \cite{Blumenhagen:2013fgp}. Moreover, there can be additional contributions at 1-loop arising from terms which a priori are of order ${\cal O}(k^4)$ after contraction of the worldsheet fields and which become of order ${\cal O}(k^2)$ only due to pinching singularities in the integration over the vertex operator positions. An example where this kind of contribution was crucial in order to get the complete kinematical structure of the Einstein-Hilbert term at 1-loop, can be found in \cite{Forger:1996vj}.}  
\be \label{1-loop-EF-action-string-1}
S_4 &=& \frac{1}{\kappa_4^2} \int d^4 x \sqrt{-g}  \left[\frac{1}{2} \left(1+ e^{2 \Phi_4} \,\delta E\right) R  
 - \sum_{i, j=0}^3  \left( G^{(0)}_{t_i t_j} + e^{2 \Phi_4} \, \overline{G}^{(1)}_{t_i t_j} \right) \, \partial_\mu t_i \partial^\mu t_j  
 \right. \nonumber \\
&& \qquad \qquad \qquad \qquad  \qquad \left. - \sum_{i, j=0}^3  \left( G^{(0)}_{c_i^{(0)} c_j^{(0)}} + 
e^{2 \Phi_4} \, \overline{G}^{(1)}_{c_i^{(0)} c_j^{(0)}}  \right) \, \partial_\mu c_i^{(0)} \partial^\mu c_j^{(0)} \right] + \ldots \,.
\ee
In \eqref{1-loop-EF-action-string-1}, we allowed for non-trivial off-diagonal metric components $\overline{G}^{(1)}_{t_i t_j}$ and $\overline{G}^{(1)}_{c_i^{(0)} c_j^{(0)}}$. A few words concerning the perturbative expansion are in order here. One might wonder about corrections to the Einstein-Hilbert term and the sigma model metric from the disk or projective plane. From the momentum expansion of the closed string 2-point functions in \cite{Hashimoto:1996bf} and in appendix A.2.\ of \cite{Lust:2004cx} it seems a priori that there are no corrections at this order (as there are no terms in the amplitudes at quadratic order in the momenta). On the other hand, in \cite{Green:2016tfs} it was conjectured that there is an $\epsilon_{10}\epsilon_{10}R^4$-term in the type I theory in 10 dimensions at disk level. Upon dimensional reduction this should lead to a disk level correction to the Einstein-Hilbert term in 4 dimensions, cf.\ \cite{Antoniadis:1997eg}. It is a very interesting question how to resolve this apparent conflict. However, we will not pursue this any further in this paper.  

After performing the Weyl rescaling 
\be
g_{\mu \nu} \rightarrow \Omega^{2} g_{\mu \nu} \,
\ee 
with $\Omega^2=\left(1+ e^{2 \Phi_4} \,\delta E\right)^{-1}$, \eqref{1-loop-EF-action-string-1} turns into the Einstein frame action. Up to 1-loop order (hence ignoring any $\left(\partial \ln \Omega \right)^2$ terms, which are of order $\mathcal{O}(e^{4 \Phi_4})$) it reads
\be \label{1-loop-EF-action-2}
S_4 &=& \frac{1}{\kappa_4^2} \int d^4 x \sqrt{-g}  \left[\frac{1}{2}  R  
 - \sum_{i, j=0}^3  \left( G^{(0)}_{t_i t_j }+ G^{(1)}_{t_i t_j} \right) \, \partial_\mu t_i \partial^\mu t_j 
 \right. \nonumber \\ 
 && \qquad \qquad \qquad \qquad  \quad \left.
 - \sum_{i, j=0}^3  \left( G^{(0)}_{c_i^{(0)} c_j^{(0)}} + 
G^{(1)}_{c_i^{(0)} c_j^{(0)}}  \right) \, \partial_\mu c_i^{(0)} \partial^\mu c_j^{(0)} \right]  + \ldots
\ee
with
\be
G^{(1)}_{t_i t_j}&=&  e^{2 \Phi_4} \,\left( \overline{G}^{(1)}_{t_i t_j} - \delta E \, G^{(0)}_{t_i t_j}\right) \,,\label{Einstein_frame_Gtt} \\
G^{(1)}_{c_i^{(0)} c_j^{(0)}} &=& e^{2 \Phi_4} \,\left(\overline{G}^{(1)}_{c_i^{(0)} c_j^{(0)}}  - \delta E \,G^{(0)}_{c_i^{(0)} c_j^{(0)}} \right) \label{Einstein_frame_Gcc} \,.
\ee

On the other hand, one could perform a Weyl rescaling $g_{\mu \nu} \rightarrow e^{-2 (\Phi_4 - \overline{\Phi}_4)}\, g_{\mu \nu}$ in \eqref{1-loop-EF-action-string-1}, which then becomes 
\be \label{1-loop-EF-action-string_frame}
S_4 &=& \frac{1}{\tilde \kappa_4^2} \int d^4 x \sqrt{-g} \Big[{1 \over 2} \left(e^{-2 \Phi_4}  + \delta E\right) R  
 + 3 \left(e^{-2 \Phi_4}  - \delta E\right)  \partial_\mu \Phi_4 \partial^\mu \Phi_4
  -  3\, \,\partial_\mu \left( \delta E \right) \partial^{\mu} \Phi_4 \\
 && - \sum_{i, j=0}^3  \left( e^{-2 \Phi_4} G^{(0)}_{t_i t_j} +  \overline{G}^{(1)}_{t_i t_j} \right) \, \partial_\mu t_i \partial^\mu t_j - \sum_{i, j=0}^3  \left( e^{-2 \Phi_4} G^{(0)}_{c_i^{(0)} c_j^{(0)}} +  
\overline{G}^{(1)}_{c_i^{(0)} c_j^{(0)}}  \right) \, \partial_\mu c_i^{(0)} \partial^\mu c_j^{(0)} \Big]  + \ldots
 \, ,  \nonumber
\ee
where we used
\be
\kappa_4 = g_s \tilde \kappa_4 = e^{\overline \Phi_4} \tilde \kappa_4\ .
\ee
The action \eqref{1-loop-EF-action-string_frame} is the conventional string-frame action, having the correct dilaton-counting for the tree and 1-loop metrics. We mention in passing that 
\be
\tilde \kappa_4^{-2} = (2 \pi \sqrt{\alpha'})^6 \kappa_{10}^{-2} = (\pi \alpha')^{-1}\ .
\ee

It is the metric of the variables $t_i$ that one has direct access to via string scattering amplitudes.\footnote{The most direct and simplest way to calculate the 1-loop corrections $\overline G^{(1)}$ and $\delta E$ in \eqref{1-loop-EF-action-string-1} would be via 2-point functions, using a procedure to relax momentum conservation which was introduced in \cite{Minahan:1987ha}. It is based on the fact that momentum conservation has a very different origin in string theory than on-shellness. Whereas the latter is required for consistency by BRST symmetry, the former only arises after integrating over the zero modes of the string coordinates and one could postpone this integration until the very end of the calculation. For type II orientifolds the procedure of \cite{Minahan:1987ha} was used to calculate the 1-loop contributions to the Einstein-Hilbert term in \cite{Antoniadis:1996vw,Antoniadis:2002tr,Kohlprath:2003pu,Epple:2004ra,Haack:2015pbv} and to scalar metrics in \cite{Berg:2005ja,Berg:2011ij,Berg:2014ama}. If one does not want to rely on relaxing momentum conservation, one would have to calculate a 4-point function of two scalars and two gravitons in order to read off the 1-loop correction to the scalar metric. Such a string 4-point function would also include wave function renormalization diagrams of the external legs and, thus, according to \cite{Antoniadis:1997eg} it would actually directly calculate the 1-loop correction to the metric in Einstein frame, i.e.\ \eqref{Einstein_frame_Gtt}. \label{2pt4pt}} However, in order to make the K\"ahler structure of the resulting metric manifest (i.e.\ the fact that the sigma model metric can be expressed as the second derivative of a K\"ahler potential), one has to use different variables. The need for changing from the string theory field variables to supergravity field variables in order to put the Lagrangian into the standard supergravity form was first discussed in the context of the heterotic string, cf.\ \cite{Witten:1985xb,Burgess:1985zz}. In our case, the kinetic terms of the tree-level action become manifestly K\"ahler when using the coordinates 
\be
T_i^{(0)} = c_i^{(0)} + i \tau_i^{(0)}\ , \label{tree-T}
\ee
where the $\tau_i^{(0)}$ are defined via \cite{Aldazabal:1998mr}
\be \label{taut}
\tau_0^{(0)} = e^{-t_0} \sqrt{t_1 t_2 t_3} \ , \quad  \tau_1^{(0)} = e^{-t_0} \sqrt{\frac{t_1}{t_2 t_3}} \ , \quad  \tau_2^{(0)} = e^{-t_0} \sqrt{\frac{t_2}{t_1 t_3}} \ , \quad  \tau_3^{(0)} = e^{-t_0} \sqrt{\frac{t_3}{t_1 t_2}}\ . 
\ee 
This can be inverted to give 
\be
e^{2 t_0} &=& \frac{1}{\sqrt{\tau_0^{(0)} \tau_1^{(0)} \tau_2^{(0)} \tau_3^{(0)}}} \ ,\label{t0-tau} \\
t_i  &=& \tau_i^{(0)}  \sqrt{\frac{\tau_0^{(0)}}{\tau_1^{(0)} \tau_2^{(0)} \tau_3^{(0)}}}\ , \qquad i \in \{ 1,2,3 \} \ . \label{t-tau0} 
\ee 
The fields \eqref{tree-T} are the dilaton and the (diagonal, untwisted) K\"ahler moduli of the tree-level supergravity action. 

As shown in appendix \ref{app:ttau}, the metric for the variables $t_i$ can be expressed through the $\tau_i^{(0)}$ via
\be \label{changettau2}
\sum_{i,j=0}^3 \left[ G^{(0)}_{t_i t_j} +  G^{(1)}_{t_i t_j} \right] \partial_\mu t_i \partial^\mu t_j = \sum_{i,j=0}^3 \left(G_{\tau_i^{(0)} \tau_j^{(0)}}^{(0)}+ G_{\tau_i^{(0)} \tau_j^{(0)}}^{(1)} \right) \partial_\mu \tau_i^{(0)} \partial^\mu \tau_j^{(0)} \,
\ee
with
\be
G_{\tau_i^{(0)}\tau_j^{(0)}}^{(0)}(\tau^{(0)}) &=& \frac{\delta_{ij}}{4 (\tau_i^{(0)})^2} \ ,\\
G_{\tau_i^{(0)}\tau_j^{(0)}}^{(1)}(\tau^{(0)}) &=& \frac{Y^{(1)}_{ij}}{\tau_i^{(0)} \tau_j^{(0)}} = \frac{\left(A^T X^{(1)} A\right)_{ij}}{\tau_i ^{(0)}\tau_j^{(0)}} \ , \label{Gtt-Y}
\ee
where the matrices $A$ and $X^{(1)}$ are given by
\be
A= \frac{1}{2} 
\begin{pmatrix}
-1 & -1 & -1 &-1 \\
 1 &  1 & -1 &-1 \\
 1 & -1 &  1 &-1 \\
 1 & -1 & -1 & 1 
\end{pmatrix}  \label{A}
\ee
and 
\be
X^{(1)} =
\begin{pmatrix}
\frac{  G^{(1)}_{t_0 t_0}}{4}, & \frac{t_1 G^{(1)}_{t_0 t_1}}{2}, & \frac{t_2 G^{(1)}_{t_0 t_2}}{2}, & \frac{t_3 G^{(1)}_{t_0 t_3}}{2} \\
\frac{t_1 G^{(1)}_{t_0 t_1}}{2}, & t_1^2 G^{(1)}_{t_1 t_1}, &  t_1 t_2 G^{(1)}_{t_1 t_2}, &  t_1 t_3 G^{(1)}_{t_1 t_3} \\
\frac{t_2 G^{(1)}_{t_0 t_2}}{2}, &  t_1 t_2 G^{(1)}_{t_1 t_2},  & t_2^2 G^{(1)}_{t_2 t_2}, & t_2 t_3 G^{(1)}_{t_2 t_3} \\
\frac{t_3 G^{(1)}_{t_0 t_3}}{2}, & t_1 t_3 G^{(1)}_{t_1 t_3},  & t_2 t_3 G^{(1)}_{t_2 t_3}, & t_3^2 G^{(1)}_{t_3 t_3}
\end{pmatrix} \,. \label{X1}
\ee
Let us also mention here that at tree-level one has \cite{Antoniadis:1996vw}
\be
G_{c_i^{(0)} c_j^{(0)}}^{(0)}(\tau^{(0)}) = G_{\tau_i^{(0)}\tau_j^{(0)}}^{(0)}(\tau^{(0)}) = \frac{\delta_{ij}}{4 (\tau_i^{(0)})^2}\ . \label{Gcc0}
\ee

The explicit form of $Y^{(1)}$ can be found in \eqref{Y00}-\eqref{Y33} of appendix \ref{app:ttau}. As an example of how the metric in $\tau$-variables looks like, let us take a closer look at $G_{\tau_3^{(0)}\tau_3^{(0)}}^{(1)}$, for instance. It is given by
\be
G_{\tau_3^{(0)}\tau_3^{(0)}}^{(1)}(\tau^{(0)}) &=& \frac{Y_{33}^{(1)}}{(\tau_3 ^{(0)})^2} = \frac{\left(A^T X^{(1)} A\right)_{33}}{(\tau_3 ^{(0)})^2} \label{Gtau3tau3}\\
&=& \frac{1}{4 (\tau_3 ^{(0)})^2}\, \left[ \frac{G_{t_0 t_0}^{(1)}}{4} + t_1 G_{t_0 t_1}^{(1)} + t_2 G_{t_0 t_2}^{(1)} -t_3 G_{t_0 t_3}^{(1)} + 
t_1^2  G_{t_1 t_1}^{(1)} +t_2^2  G_{t_2 t_2}^{(1)}+t_3^2  G_{t_3 t_3}^{(1)}  \right. \nonumber \\ 
&& \left. \qquad \qquad \qquad  + 2 t_1 t_2  G_{t_1 t_2}^{(1)} - 2 t_1 t_3  G_{t_1 t_3}^{(1)} - 2 t_2 t_3  G_{t_2 t_3}^{(1)} \right] \\
&=&  \frac{e^{2 \Phi_4}}{4 (\tau_3^{(0)})^2}\left[\frac{{\overline G}^{(1)}_{t_0 t_0}}{4} + \left(t_1 \overline G_{t_0 t_1}^{(1)}+t_2 \overline G_{t_0 t_2}^{(1)}-t_3 \overline G_{t_0 t_3}^{(1)} \right) + \sum_{i=1}^3 \left(t_i^2 \overline G_{t_i t_i}^{(1)}\right)
\right. \nonumber \\
&&\qquad  \qquad \left. + 2 \left(t_1 t_2 \overline G_{t_1 t_2}^{(1)}-t_1 t_3 \overline G_{t_1 t_3}^{(1)}-t_2 t_3 \overline G_{t_2 t_3}^{(1)} \right) - \delta E 
\right] \,. \label{Gtt_33}
\ee
In the third equality we used \eqref{Einstein_frame_Gtt}. The final result has to be understood as a function of $\tau_i^{(0)}$ (using \eqref{t0-tau}  and \eqref{t-tau0}). Equation \eqref{Gtt_33} expresses the metric component $G_{\tau_3^{(0)}\tau_3^{(0)}}^{(1)}$ in terms of the metric components $\overline G_{t_i t_j}^{(1)}$ and the correction to the Einstein-Hilbert term $\delta E$, all of which are directly calculable via string 2-point functions, cf.\ footnote \ref{2pt4pt}. 

Some comments are in order here. In \cite{Berg:2005ja,Berg:2014ama} a different strategy for calculating the 1-loop corrections to the moduli metric was followed. To understand this, we first observe that the tree-level fields $\tau_i^{(0)}$ can also be expressed through the 10-dimensional dilaton $\Phi_{10}$. Using  $e^{\Phi_{10}} = e^{t_0} \sqrt{t_1 t_2 t_3}$ one easily verifies that \eqref{taut} can be written as
\be \label{tautphi10}
\tau_0^{(0)} = e^{-\Phi_{10}} t_1 t_2 t_3 \ , \quad  \tau_1^{(0)} = e^{-\Phi_{10}} t_1 \ , \quad  \tau_2^{(0)} = e^{-\Phi_{10}} t_2 \ , \quad  \tau_3^{(0)} = e^{-\Phi_{10}} t_3\ .
\ee
In \cite{Berg:2005ja,Berg:2014ama} the value for $\Phi_{10}$ was {\it fixed}. In that case the vertex operator for $\tau_i^{(0)}$ is the same as the one for $t_i$, up to a constant rescaling by $e^{-\Phi_{10}}$. It is these vertex operators for $\tau_i^{(0)}$ that were used in \cite{Berg:2005ja,Berg:2014ama} to calculate the metric for the $\tau_i^{(0)}$. For the example of $G_{\tau_3^{(0)}\tau_3^{(0)}}^{(1)}$ (given in \eqref{Gtau3tau3}-\eqref{Gtt_33}), effectively this amounts to calculating $\overline G_{t_3 t_3}^{(1)}$ instead. However, note that all the terms in the square bracket receive the same moduli dependence from a given orbifold sector and a given worldsheet topology (i.e.\ annulus ${\cal A}$, M\"obius ${\cal M}$, Klein bottle ${\cal K}$ or torus ${\cal T}$).\footnote{This will become clear in sections \ref{N=1} and \ref{sec.N=2}, cf.\ \eqref{tscalingN1} and \eqref{tscaling} together with the notation of \eqref{m1-1}.} Thus, the procedure of \cite{Berg:2005ja,Berg:2014ama} allows one to calculate the right moduli dependence for the metric of the $\tau_i^{(0)}$ but one can not calculate the correct coefficients in this way. In \cite{Berg:2005ja} the coefficients were left undetermined and only in the ${\cal N} = 2$ case of a $\mathbb{T}^2 \times {\rm K}3$-compactification (cf.\ appendix \ref{sec:t2k3} below) the explicit coefficient of the K\"ahler potential could be obtained indirectly using the higher amount of supersymmetry (cf.\ appendix D in \cite{Berg:2005ja}). It is one goal of the present paper to present formulas indicating which combination of string 2-point functions one would have to calculate in order to fix the explicit coefficients in the 1-loop K\"ahler potential (and the 1-loop field redefinitions). 

Moreover, note that we are always using the language of type IIB orientifolds with D9/D5-branes. However, our final results for the 1-loop field redefinitions and the correction to the K\"ahler potential should also be valid for  type IIB orientifolds with D3/D7-branes, using $\tau_0^{(0)} = e^{-\Phi_{10}}$ and $\tau_i^{(0)} = e^{-\Phi_{10}} \frac{t_1 t_2 t_3}{t_i}$ instead of \eqref{tautphi10}.


\section{Field redefinition and K\"ahler potential at 1-loop level: General framework} \label{redef}

In this section we would like to discuss the general strategy to obtain the 1-loop field redefinition and the 1-loop correction to the K\"ahler potential, once the corrections to the moduli metric and the Einstein-Hilbert term have been calculated, cf.\ \eqref{1-loop-EF-action-string-1}.\footnote{This strategy and the results for the $\mathbb{Z}_6'$ orientifold of section \ref{applicationZ6} below were partly already summarized in \cite{Haack:2017vko}.} As mentioned before, at tree-level the moduli metric is K\"ahler with the K\"ahler coordinates \eqref{tree-T} and K\"ahler potential
\be \label{K0T0}
K^{(0)}(T^{(0)}, \bar{T}^{(0)}) = - \sum_{i=0}^3 \ln \left( T_i^{(0)} - \bar T_i^{(0)} \right)\ .
\ee
In principle one could now proceed to calculate the 1-loop corrections to the moduli metric using the well known vertex operators for $t_i$ and $c_i^{(0)}$ and express these (after a Weyl-rescaling to the Einstein frame, cf.\ \eqref{1-loop-EF-action-2}) in terms of $c_i^{(0)}$ and $\tau_i^{(0)}$, using \eqref{t0-tau} and \eqref{t-tau0}, cf.\ \eqref{changettau2}.\footnote{Note that for $c_0^{(0)}$ one would first have to calculate the kinetic term and a possible Chern-Simons term  for $C_{\mu \nu}$ and then dualize. Alternatively, one may be able to use the vertex operator for the RR 6-form $C_6$ with only internal indices.} The result for the kinetic terms would then look like
\be \label{loopcorrectedL}
{\cal L}_{\rm kin} \sim - \sum_{i,j=0}^3 \left( G_{c_i^{(0)} c_j^{(0)}} (\tau^{(0)}) \partial_\mu c_i^{(0)} \partial^\mu c_j^{(0)} + G_{\tau_i^{(0)} \tau_j^{(0)}} (\tau^{(0)}) \partial_\mu \tau_i^{(0)} \partial^\mu \tau_j^{(0)} \right) + \ldots \ ,
\ee
where $G = G^{(0)} + G^{(1)}$. The metric components have to be independent of $c^{(0)}$ due to its perturbative shift symmetry.\footnote{We do not consider any non-perturbative corrections, neither on the world-sheet nor in space-time, which might break this shift symmetry to a discrete subgroup.} Moreover, (parity even) amplitudes with a single RR-vertex operator vanish and, thus, there is no mixed term of the form $G_{c_i^{(0)} \tau_j^{(0)}}$. In general, the moduli metric in \eqref{loopcorrectedL} will not be K\"ahler anymore for the coordinates \eqref{tree-T} and, in that case, one can not directly read off the corrections to the K\"ahler potential from the metric. Rather, one has to find 1-loop corrected variables 
\be \label{Tj}
T_j =  c_j + i \tau_j = c_j^{(0)} + c_j^{(1)}(c^{(0)}, \tau^{(0)}) + i \left( \tau_j^{(0)} + \tau_j^{(1)}(c^{(0)}, \tau^{(0)})  \right) \ ,
\ee
which are not holomorphically related to \eqref{tree-T} and for which the metric in  \eqref{loopcorrectedL} becomes K\"ahler, i.e.
\be
{\cal L}_{\rm kin} \sim - \sum_{i,j=0}^3 \frac{\partial^2 K}{\partial T_i \partial \bar T_j}\, \partial_\mu T_i \partial^\mu \bar T_j + \ldots \ ,
\ee
where the K\"ahler potential $K$ includes a 1-loop correction $K^{(1)}$, i.e.\ 
\be \label{fullK}
K(T, \bar{T})=K^{(0)}(T, \bar{T}) + K^{(1)}(T, \bar{T})\ . 
\ee
Note that $K^{(0)}(T, \bar{T})$ in \eqref{fullK} takes the same form as in \eqref{K0T0}, but with $T^{(0)}_i$ replaced by the corrected variables $T_i$. A priori, the 1-loop corrections to $c_j^{(0)}$ and $\tau_j^{(0)}$ might depend on both the $c^{(0)}$s and the $\tau^{(0)}$s, but due to the shift symmetry of the $c^{(0)}$s one can actually restrict the ansatz for the field redefinition to 
\begin{empheq}[box=\fbox]{align}
c_j = c_j^{(0)}\quad , \quad  \tau_j = \tau_j^{(0)} + \tau_j^{(1)}(\tau^{(0)}) \ . \label{c_tau_redef} 
\end{empheq}
Let us go through the argument for this. It contains two steps. In a first step, we argue that one can choose K\"ahler coordinates such that the $c^{(1)}$s and $\tau^{(1)}$s do not depend on the $c^{(0)}$s so that the corrected $T_j$ still fulfill
\be \label{Ttransform}
T_j \stackrel{c_k^{(0)} \rightarrow c_k^{(0)} + a_k}{\longrightarrow} T_j + a_j
\ee
like at tree-level. Assume we found some K\"ahler coordinates \eqref{Tj} which do not fulfill this. Then under infinitesimal shifts $c_k^{(0)} \rightarrow c_k^{(0)} + a_k$, which should correspond to symmetries of the K\"ahler manifold, these K\"ahler variables would transform according to 
\be
\delta T_j = a_j + \sum_k \left( \frac{\partial c_j^{(1)}}{\partial c_k^{(0)}} + i \frac{\partial \tau_j^{(1)}}{\partial c_k^{(0)}} \right) a_k \equiv  a_j + \sum_k f_{jk} (T) a_k\ ,
\ee
where in the last step we defined the functions $f_{jk} (T)$. These have to depend {\it holomorphically} on the K\"ahler variables $T$, given that symmetries of a K\"ahler manifold are described by holomorphic Killing vectors in order to preserve the complex structure (cf.\ section 13.4 in \cite{Freedman:2012zz}, for instance). Note that the functions $f_{jk}$ are suppressed by a factor $g_s^2$, as they arise at 1-loop level. One could now define new variables
\be
T_j' = T_j - \sum_k \int^{T_k} d \tilde T_k \ f_{jk}(\tilde T) = c_j^{(0)} + (c_j')^{(1)} + i \left( \tau_j^{(0)} + (\tau_j')^{(1)}  \right)\ ,
\ee
which are also valid K\"ahler coordinates, as the coordinate change is holomorphic. However, the new coordinates transform under infinitesimal shifts $c_k^{(0)} \rightarrow c_k^{(0)} + a_k$ according to 
\be
\delta T_j' = \delta T_j - \sum_k f_{jk} (T) \delta T_k = a_j + {\cal O}(g_s^4)\ ,
\ee
i.e.\ $(c')^{(1)}$s and $(\tau')^{(1)}$s do not depend on the $c^{(0)}$s. Thus, we can now focus on field redefinitions of the form 
\be \label{Tnoc0}
T_j =  c_j + i \tau_j = c_j^{(0)} + c_j^{(1)}(\tau^{(0)}) + i ( \tau_j^{(0)} + \tau_j^{(1)}(\tau^{(0)}))\ ,
\ee
for which \eqref{Ttransform} is still satisfied.  

In a second step we show that the $c^{(1)}$s can actually be set to zero. Due to the symmetry of the theory under shifts 
\eqref{Ttransform}, the K\"ahler potential can only depend on the imaginary parts of the $T$s, i.e.\ 
\be
K(T, \bar T) = K(\tau)\ .
\ee
This implies 
\be
\frac{\partial^2 K}{\partial T_i \partial \bar T_j}\, \partial_\mu T_i \partial^\mu \bar T_j = \frac14 \frac{\partial^2 K}{\partial \tau_i \partial \tau_j} \left( \partial_\mu c_i \partial^\mu c_j +  \partial_\mu \tau_i \partial^\mu \tau_j \right)\ 
\ee
or in other words
\bea
G_{c_i \tau_j} &=& 0\ , \label{Gctau} \\
G_{\tau_i \tau_j} &=& \frac14 \frac{\partial^2 K}{\partial \tau_i \partial \tau_j} = G_{c_i c_j}\ .  \label{Gtautaucc} 
\eea
On the other hand, expressing the Lagrangian \eqref{loopcorrectedL} in terms of the new variables by substituting $c_i^{(0)} = c_i - c_i^{(1)}(\tau^{(0)})$ and $\tau_i^{(0)} = \tau_i - \tau_i^{(1)}(\tau^{(0)})$ into \eqref{loopcorrectedL}, we obtain
\be
{\cal L}_{\rm {kin}} \sim - \sum \left( G_{c_i c_j} \partial_{\mu } c_i \partial^{\mu } c_j + 2 G_{c_i \tau_j} \partial_{\mu } c_i \partial^{\mu } \tau_j + G_{\tau_i \tau_j} \partial_{\mu } \tau_i \partial^{\mu } \tau_j \right) \ ,
\ee
where\footnote{Note that in $G^{(1)}$ and in terms involving $c^{(1)}$ or $\tau^{(1)}$ we can interchange $\tau^{(0)}$ and $\tau$ to 1-loop order. The same holds true for the 1-loop correction to the K\"ahler potential $K^{(1)}$. \label{interchange}} 
\bea
G_{c_i \tau_j} &=& - \sum_k \left[G_{c_i^{(0)} c_k^{(0)}}^{(0)} (\tau)\, \partial_{\tau_j} \left(c_k^{(1)}\right) \right] = -\frac{ \partial_{\tau_j} \left(c_i^{(1)}\right)}{4 \tau_i^2} \ , \label{mixing} \\
G_{\tau_i \tau_j} &=& G_{\tau_i^{(0)} \tau_j^{(0)}}(\tau) - \sum_k \tau_k^{(1)} \frac{\partial}{\partial  \tau_k}\left(G_{\tau_i^{(0)} \tau_j^{(0)}}^{(0)}(\tau) \right) 
- \sum_k\left[G_{\tau_i^{(0)} \tau_k^{(0)}}^{(0)}(\tau)\, \partial_{\tau_j} \left(\tau_k^{(1)}\right) +(i \leftrightarrow j) \right]  \nonumber \\
&=& \frac{\delta_{ij}}{4 \tau_i^2} + G_{\tau_i^{(0)} \tau_j^{(0)}}^{(1)}(\tau) +  \frac{\tau_i^{(1)}}{2 \tau_i^3} \delta_{ij}
- \frac{1}{4} \left[ \frac{\partial_{\tau_j} \left(\tau_i^{(1)}\right)}{\tau_i^2} + \frac{\partial_{\tau_i} \left(\tau_j^{(1)}\right)}{\tau_j^2} \right] \ , \label{appG_tau_tau} \\
G_{c_i c_j} &=& G_{c_i^{(0)} c_j^{(0)}}(\tau) - \sum_k \tau_k^{(1)} \frac{\partial}{\partial  \tau_k}\left(G_{c_i^{(0)} c_j^{(0)}}^{(0)}(\tau) \right) = \frac{\delta_{ij}}{4 \tau_i^2} + G_{c_i^{(0)} c_j^{(0)}}^{(1)}(\tau) +  \frac{\tau_i^{(1)}}{2 \tau_i^3} \delta_{ij} \ .  \label{appG_c_c}
\eea
Comparing \eqref{mixing} with \eqref{Gctau} we see that the $c^{(1)}$s can also not depend on the $\tau$s and, thus, can be chosen to vanish (a constant shift amounts to a holomorphic field redefinition). This completes the proof that we can restrict the field redefinitions to the form \eqref{c_tau_redef}. 

We can now use \eqref{Gtautaucc}, \eqref{appG_tau_tau} and \eqref{appG_c_c} in order to derive conditions on the 1-loop corrections to the field variables and the K\"ahler potential, $\tau^{(1)}$ and $K^{(1)}$. Concretely, the equations for the $\tau^{(1)}$s are
\begin{empheq}[box=\fbox]{align}
\frac{\partial_{\tau_j} \left(\tau_i^{(1)}\right)}{2 \tau_i^3}  & = \partial_{\tau_i}\left(G_{c_i^{(0)} c_j^{(0)}}^{(1)}(\tau)\right) - \partial_{\tau_j}\left(G_{c_i^{(0)} c_i^{(0)}}^{(1)}(\tau)\right)  \qquad {\rm for} \quad i\neq j \ , \label{eq-ij}   \\
\frac{\partial_{\tau_i} \left(\tau_i^{(1)}\right)}{ 2 \tau_i^2}  & = G_{\tau_i^{(0)} \tau_i^{(0)}}^{(1)}(\tau) -  G_{c_i^{(0)} c_i^{(0)}}^{(1)}(\tau)  \label{eq-ii}
\end{empheq}
and the equation for $K^{(1)}$ is
\be
\frac{1}{4} \frac{\partial^2 K^{(1)}(\tau)}{\partial \tau_i \partial \tau_j} = G_{c_i^{(0)} c_j^{(0)}}^{(1)}(\tau) +  \frac{\tau_i^{(1)}}{2 \tau_i^3} \delta_{ij} \label{K1eq}
\ee
or
\begin{empheq}[box=\fbox]{align}
\frac{1}{4} \frac{\partial^2 K^{(1)}(\tau)}{\partial \tau_i \partial \tau_j} & = G_{c_i^{(0)} c_j^{(0)}}^{(1)}(\tau)  \qquad {\rm for} \quad i\neq j \ ,  \label{K1ij} \\
\frac{1}{4} \frac{\partial^2 K^{(1)}(\tau)}{\partial \tau_i \partial \tau_i} & = G_{c_i^{(0)} c_i^{(0)}}^{(1)}(\tau) +  \frac{\tau_i^{(1)}}{2 \tau_i^3}\ . \label{K1ii}
\end{empheq}
Note that the right hand sides of \eqref{eq-ij} and \eqref{eq-ii}, which source the field redefinitions, measure the ``non-K\"ahlerness" of the 1-loop metric, expressed in terms of $\tau_i^{(0)}$ and $c_i^{(0)}$. 

Eq.\ \eqref{K1eq} follows directly from \eqref{Gtautaucc} and \eqref{appG_c_c} and eq.\ \eqref{eq-ii} is a consequence of $G_{c_i c_j} = G_{\tau_i \tau_j}$ and \eqref{appG_tau_tau} and \eqref{appG_c_c} which imply 
\be
\frac{1}{4} \left[\frac{\partial_{\tau_j} \left(\tau_i^{(1)}\right)}{\tau_i^2} + \frac{\partial_{\tau_i} \left(\tau_j^{(1)}\right)}{\tau_j^2} \right]
=G_{\tau_i^{(0)} \tau_j^{(0)}}^{(1)}(\tau) -  G_{c_i^{(0)} c_j^{(0)}}^{(1)}(\tau)\ . \label{eq1}
\ee
Eq.\ \eqref{eq-ii} follows when choosing $i = j$. Finally, eq.\ \eqref{eq-ij} follows easily from \eqref{K1eq} and the identity
\be
\frac{\partial}{\partial \tau_j}\left(\frac{\partial^2 K^{(1)}(\tau)}{\partial \tau_i \partial \tau_i}\right) = 
\frac{\partial}{\partial \tau_i}\left(\frac{\partial^2 K^{(1)}(\tau)}{\partial \tau_i \partial \tau_j}\right)\ . 
\ee

To sum up, equations \eqref{eq-ij} and \eqref{eq-ii} should be solved to find the field redefinitions $\tau_i^{(1)}(\tau)$ and the 1-loop K\"ahler potential $K^{(1)}$ is then obtained by integrating \eqref{K1eq} once the $\tau_i^{(1)}(\tau)$ are known. One comment is in order here. Naively, it appears as if one needs to determine the correction to the metric of the axions in order to obtain the field redefinitions and the 1-loop correction to the K\"ahler potential. A knowledge of the 1-loop correction to the metric of the $\tau_i^{(0)}$ alone seems not sufficient. However, we will discuss in a moment that the metric components are not all independent but rather have to obey some consistency conditions. Using a particular ansatz for the form of the 1-loop corrections to the moduli metric, we will see in section \ref{N=1} and \ref{sec.N=2} that knowledge of the 1-loop correction to the metric of the $\tau_i^{(0)}$ is actually sufficient.

In order for the solutions of the above equations \eqref{eq-ij}-\eqref{K1eq} to exist, there are consistency conditions on the loop corrections $G^{(1)}$. Concretely, we find the following independent conditions:
\begin{empheq}[box=\fbox]{align}
& \tau_i \left[\partial_{\tau_i}  G_{c_i^{(0)} c_j^{(0)}}^{(1)}(\tau) - \partial_{\tau_j} \, G_{c_i^{(0)} c_i^{(0)}}^{(1)}(\tau) \right] + (i \leftrightarrow j) 
= 2 \left[G_{\tau_i^{(0)} \tau_j^{(0)}}^{(1)}(\tau) -G_{c_i^{(0)} c_j^{(0)}}^{(1)}(\tau)\right]  \quad (i\neq j)\ , \label{cons-cond1} \\
& \partial_{\tau_k} G_{c_i^{(0)} c_j^{(0)}}^{(1)}(\tau)
= \partial_{\tau_i} G_{c_j^{(0)} c_k^{(0)}}^{(1)}(\tau)=\partial_{\tau_j} G_{c_k^{(0)} c_i^{(0)}}^{(1)}(\tau) \qquad  (i\neq j \neq k)\ ,  \label{cons-cond2} \\
& \tau_i^2 \frac{\partial}{\partial \tau_j}\left[G_{c_i^{(0)} c_i^{(0)}}^{(1)}(\tau) - G_{\tau_i^{(0)} \tau_i^{(0)}}^{(1)}(\tau)\right]
= \frac{\partial}{\partial \tau_i} \left[\tau_i^3 \left( \partial_{\tau_j}\,G_{c_i^{(0)} c_i^{(0)}}^{(1)}(\tau) - \partial_{\tau_i}\, G_{c_i^{(0)} c_j^{(0)}}^{(1)}(\tau) \right)\right] \quad (i\neq j) \ .  \label{cons-cond3}
\end{empheq}
Eq.\ \eqref{cons-cond1} follows from \eqref{eq-ij} (plus the one with $(i\leftrightarrow j)$) and \eqref{eq1}, eq.\ 
\eqref{cons-cond2} can be inferred from \eqref{K1eq} with $i\neq j$ and from the invariance of $\frac{\partial^3\left(K^{(1)}(\tau)\right)}{\partial \tau_i \partial \tau_j \partial \tau_k}$ under permutation of the derivatives, and eq.\ \eqref{cons-cond3} arises from \eqref{eq-ij}-\eqref{eq-ii} with properly taken derivatives and using $\partial_{\tau_i}\partial_{\tau_j} \left(\tau_i^{(1)}\right)=\partial_{\tau_j}\partial_{\tau_i} \left(\tau_i^{(1)}\right)$. The 1-loop metric corrections $G^{(1)}$ must satisfy the consistency conditions \eqref{cons-cond1}-\eqref{cons-cond3} otherwise one can not express the metric via a K\"ahler potential with appropriate K\"ahler coordinates at 1-loop order.    


\section{Concrete computations}
\label{concretecomps}

In order to proceed further, we now have to make an ansatz for the field dependence of the 1-loop corrected moduli metric. We will do so in the following sections, separately for the contributions of the $\mathcal{N}=1$ and $\mathcal{N}=2$ sectors ($\mathcal{N}=4$ sectors do not contribute due to the high supersymmetry). The final results for the field redefinitions and 1-loop correction to the K\"ahler potential will then be the sum of the contributions from the $\mathcal{N}=1$ and $\mathcal{N}=2$ sectors, i.e. 
\be 
\tau_i^{(1)}= \tau_i^{(1)}\Big|_{\mathcal{N}=1}+\tau_i^{(1)} \Big|_{\mathcal{N}=2} 
\ee 
and 
\be K^{(1)} = K^{(1)}_{\mathcal{N}=1}+K^{(1)}_{\mathcal{N}=2} \ .
\ee


\subsection{$\mathcal{N}=1$ sectors}  \label{N=1}

$\mathcal{N}=1$ sectors are all that one needs for odd $N$ orientifold models, while even $N$ orientifold models also contain $\mathcal{N}=2$ sectors in addition to $\mathcal{N}=1$ sectors. Those will be discussed in the next section.  
  
In $\mathcal{N}=1$ sectors, the moduli dependence of 1-loop corrections to the metric (in Einstein frame) is simple and given by
\begin{empheq}[box=\fbox]{align}
G_{\tau_i^{(0)} \tau_j^{(0)}}^{(1)}(\tau^{(0)}) &= \alpha_{ij} \frac{e^{2\Phi_4}}{\tau_i^{(0)} \tau_j^{(0)}} =   
\frac{\alpha_{ij}}{\tau_i^{(0)} \tau_j^{(0)} \sqrt{\tau_0^{(0)} \tau_1^{(0)} \tau_2^{(0)} \tau_3^{(0)}}}\,, \label{Gtt_N=1} \\
G_{c_i^{(0)} c_j^{(0)}}^{(1)}(\tau^{(0)}) &= \beta_{ij} \frac{e^{2\Phi_4}}{\tau_i^{(0)} \tau_j^{(0)}} = \frac{\beta_{ij}}{\tau_i^{(0)} \tau_j^{(0)} 
\sqrt{\tau_0^{(0)} \tau_1^{(0)} \tau_2^{(0)} \tau_3^{(0)}}}\,, \label{Gcc_N=1}
\end{empheq}
where we used $e^{2 \Phi_4}=e^{2 t_0}=(\tau_0^{(0)} \tau_1^{(0)} \tau_2^{(0)} \tau_3^{(0)})^{-1/2}$ (see \eqref{t0-tau}). Here $\alpha_{ij}$ and $\beta_{ij}$ (obviously symmetric under exchange of the indicies) are moduli-independent constants, which can be fixed by calculating 1-loop amplitudes for a specific orientifold model (hence these constants are model dependent), cf.\ \cite{Berg:2014ama} for an example. 

The moduli dependence of \eqref{Gtt_N=1} and \eqref{Gcc_N=1} can be easily understood. It comes entirely through the loop counting factor involving the dilaton and through the normalization factors of the vertex operators (which is the same as at tree-level). Let us first consider \eqref{Gtt_N=1} in order to discuss the $\tau^{(0)}$-dependence a bit more in detail. In string theory one would naturally use the vertex operators for the $t_i$ (with $i \in \{ 0,1,2,3 \}$) and the graviton in order to calculate $\overline G^{(1)}_{t_i t_j}$ and $\delta E$, cf.\ \eqref{1-loop-EF-action-string-1}. For the ${\cal N}=1$ sector contributions to these quantities, the scaling with the volume moduli $t_i$ (for $i \in \{ 1,2,3 \}$) arises solely from the normalization of the vertex operators. The vertex operators of the graviton and of the 4-dimensional dilaton $t_0$ are independent of any volume moduli, cf.\ \eqref{vertexhD}, whereas the vertex operators for the $t_i$ (with $i \in \{ 1,2,3 \}$) are proportional to $t_i^{-1}$, cf.\ equation (3.30) in \cite{Lust:2004cx} (note that their ${\rm Im} T$ corresponds to our $t$). Thus, we infer the scaling
\be \label{tscalingN1}
\delta E = {\rm const.} \ , \quad \overline{G}^{(1)}_{t_i t_j} \sim \frac{1}{t_i t_j}  \ , \quad  \overline{G}^{(1)}_{t_0 t_i} \sim \frac{1}{t_i} \ , \quad \overline{G}^{(1)}_{t_0 t_0} = {\rm const.}  \ , \quad i,j \in \{ 1,2,3 \}\ .
\ee 
From this scaling behavior, the $\tau^{(0)}$-dependence of \eqref{Gtt_N=1} can be inferred by using \eqref{Einstein_frame_Gtt} and changing the coordinates from $t$ to $\tau$, cf.\ \eqref{Gtt-Y}-\eqref{X1}. In  \eqref{Gcc_N=1} the factor $(\tau_i^{(0)} \tau_j^{(0)})^{-1}$ arises due to the normalization of the vertex operators for $c_i^{(0)}$. This normalization is fixed by supersymmetry and has to be chosen such that it reproduces the tree-level metric \eqref{Gcc0}. Namely, given that the metric \eqref{Gcc0} could be calculated by a sphere amplitude of two axions and a graviton, we see that the vertex operators for $c_i^{(0)}$ have to contain a factor of $(\tau_i^{(0)})^{-1}$. 

We now substitute \eqref{Gtt_N=1}-\eqref{Gcc_N=1} (with the replacement $\tau_i^{(0)} \to \tau_i$ in the arguments, which is allowed to 1-loop order) into the equations given in the previous section 
to find the field redefinitions $\tau_i^{(1)}$ and K\"ahler potential $K^{(1)}$.  Moreover, using the consistency conditions \eqref{cons-cond1}-\eqref{cons-cond3}, we note that the $\alpha$s and the $\beta$s are not independent. 

First, condition \eqref{cons-cond2} gives
\be
\beta_{ij}=\beta_{jk}=\beta_{ik} \equiv \beta \qquad \text{for } \quad i\neq j\neq k \,. \label{beta-cond}
\ee 
Using this, conditions \eqref{cons-cond3} and \eqref{cons-cond1} give
\be
\alpha_{ii}= \frac{1}{2}(\beta_{ii}+3 \beta)  \label{ab-cond3}
\ee
and
\be
\alpha_{ij}= \frac{1}{4} (\beta_{ii}+\beta_{jj})- \frac{\beta}{2}  \qquad \text{for } \quad i\neq j\, , \label{ab-cond1}
\ee
respectively. The above non-trivial three conditions must be fulfilled in order for K\"ahler coordinates to exist. 
Relations \eqref{ab-cond3}-\eqref{ab-cond1} can be inverted to give
\be
\beta &=& \frac{1}{4} \,(\alpha_{ii}+\alpha_{jj}- 2\,\alpha_{ij}) \,,  \label{ba-1} \\
\beta_{ii} &=& \frac{1}{4} \, (5\, \alpha_{ii}-3 \,\alpha_{jj}+ 6\, \alpha_{ij}) \label{ba-2}
\ee 
with $j \neq i$. Since \eqref{ba-1} holds for any $i\neq j$, it implies 
\be
&& \alpha_{11}+\alpha_{22} - 2 \alpha_{12} = \alpha_{11}+\alpha_{33} - 2 \alpha_{13} 
=\alpha_{00}+\alpha_{11} - 2 \alpha_{01}=  \nonumber \\
&&= \alpha_{00}+\alpha_{22} - 2 \alpha_{02} = \alpha_{00}+\alpha_{33} - 2 \alpha_{03}
=\alpha_{22}+\alpha_{33} - 2 \alpha_{23}. \label{alphas-N=1}
\ee 
Note that \eqref{beta-cond}-\eqref{ba-2} lead to non-trivial predictions of relations among the $\mathcal{N}=1$ sector contributions to NSNS and RR 1-loop 2-point amplitudes. As anticipated in the last section, \eqref{ba-1} and \eqref{ba-2} show that indeed the consistency conditions are strong enough to fix the 1-loop corrections to the metric of the axions in terms of the corresponding 1-loop corrections to the metric of the $\tau_i^{(0)}$, at least for $\mathcal{N}=1$ sectors (we will see a similar result for the $\mathcal{N}=2$ sectors in the next section).

Now, using \eqref{Gtt_N=1}-\eqref{Gcc_N=1}, we can find the 1-loop field redefinitions $\tau_i^{(1)}$ and the correction to the K\"ahler metric/potential from \eqref{eq-ij}-\eqref{eq-ii} and \eqref{K1ij}-\eqref{K1ii}, respectively. We obtain
\begin{empheq}[box=\fbox]{align}
\left. \tau_i^{(1)}\right|_{\mathcal{N}=1}= \frac{4(\alpha_{ii}-\beta_{ii})\, \tau_i}{\sqrt{\tau_0 \tau_1 \tau_2 \tau_3}} = 4(\alpha_{ii}-\beta_{ii})\, \tau_i\, e^{2 t_0} =  (-\alpha_{ii} + 3 \alpha_{jj} - 6 \alpha_{ij})\, \tau_i\, e^{2 t_0} \qquad (j \neq i) \,, \label{tau^1_N=1}
\end{empheq}
where in the second equality we used \eqref{t0-tau} (interchanging $\tau$ with $\tau^{(0)}$, cf.\ footnote \ref{interchange}) and in the last equality we used \eqref{ba-2}. The K\"ahler metric \eqref{K1ij}-\eqref{K1ii} reads
\be
\frac{\partial^2 K^{(1)}(\tau)}{\partial \tau_i \partial \tau_j} &=& 
     \frac{4 \beta}{\tau_i\tau_j \sqrt{\tau_0 \tau_1 \tau_2 \tau_3}}        \qquad \qquad \qquad \text{if } i\neq j  \,, \label{K1_N=1}  \\ 
  \frac{\partial^2 K^{(1)}(\tau)}{\partial \tau_i \partial \tau_i} &=& \frac{4 (2 \alpha_{ii}-\beta_{ii}) }{\tau_i^2 \sqrt{\tau_0 \tau_1 \tau_2 \tau_3}}  
  =  \frac{ 12 \beta }{\tau_i^2 \sqrt{\tau_0 \tau_1 \tau_2 \tau_3}}  
\label{K2_N=1}\,. 
\ee
In the last equality we used \eqref{ab-cond3}. One can check that the above expressions fulfill integrability conditions, so that we can integrate them to obtain the 1-loop correction to the K\"ahler potential from $\mathcal{N}=1$ sectors as 
\begin{empheq}[box=\fbox]{align}
K^{(1)}_{\mathcal{N}=1}= \frac{16 \beta}{\sqrt{\tau_0 \tau_1 \tau_2 \tau_3}} = \frac{4 (\alpha_{ii}+\alpha_{jj}- 2\,\alpha_{ij})}{\sqrt{\tau_0 \tau_1 \tau_2 \tau_3}} \qquad (j \neq i) \,, \label{K_N=1}
\end{empheq}
where we used \eqref{ba-1} in the last equality. It is easy to see that the K\"ahler potential above reproduces the metric in \eqref{K1_N=1} and \eqref{K2_N=1}. Note that it is possible to express both $\tau_i^{(1)}$ and $K^{(1)}_{\mathcal{N}=1}$ entirely in terms of $\alpha$s (i.e. $G_{\tau_i^{(0)} \tau_j^{(0)}}^{(1)}$) using \eqref{ba-1}-\eqref{ba-2}, without invoking $\beta$s (i.e. $G_{c_i^{(0)} c_j^{(0)}}^{(1)}$). 

We would now like to express the results \eqref{tau^1_N=1} and \eqref{K_N=1} in terms of the quantities that one actually calculates with string amplitudes, i.e.\ $\delta E$ and the components of $\overline G^{(1)}$, cf.\ \eqref{1-loop-EF-action-string-1}. To this end, let us first investigate what \eqref{alphas-N=1} entails. 
From \eqref{Gtt-Y} and \eqref{Gtt_N=1}, we have
\be
\alpha_{i j} = e^{-2 t_0} Y^{(1)}_{ij}  = e^{-2 t_0} \left(A^{T} X^{(1)} A\right)_{ij} \,. \label{alpha_Y_X}
\ee
Here $X^{(1)}$ and $A$ are given by \eqref{X1} and \eqref{A}, respectively, while $Y^{(1)}_{ij}$ is given by \eqref{Y00}-\eqref{Y33}. Inserting \eqref{Y00}-\eqref{Y33} into \eqref{alpha_Y_X}, constraints \eqref{alphas-N=1} can be solved (using \eqref{Y0-1}-\eqref{Y1-2}) to give
\be
t_1^2 G_{t_1t_1}^{(1)}= t_2^2 G_{t_2 t_2}^{(1)} = t_3^2 G_{t_3 t_3}^{(1)}\,, \qquad G_{t_1t_2}^{(1)}=  G_{t_1 t_3}^{(1)} = G_{t_2 t_3}^{(1)}=0\ ,   \label{relat-Gtt}
\ee
which implies (from \eqref{Einstein_frame_Gtt})
\be
t_1^2 \overline{G}_{t_1t_1}^{(1)}= t_2^2 \overline{G}_{t_2 t_2}^{(1)} = t_3^2 \overline{G}_{t_3 t_3}^{(1)} \, ,
\qquad \overline{G}_{t_1t_2}^{(1)}=  \overline{G}_{t_1 t_3}^{(1)} = \overline{G}_{t_2 t_3}^{(1)}=0  \ . \label{relat-tilde-Gtt}
\ee
These imply non-trivial relations between 2-point amplitudes of volume moduli of different tori which have to hold for $\mathcal{N}=1$ sectors. Then from \eqref{ba-1}, \eqref{alpha_Y_X}, \eqref{Y0-1}-\eqref{Y1-2} and \eqref{Einstein_frame_Gtt} we have 
\be
\beta= \frac{e^{-2 t_0}\, t_1^2\, G_{t_1 t_1}^{(1)}}{2} = \frac{1}{8} \left. \left(4 t_1^2\, \overline{G}_{t_1 t_1}^{(1)}- \delta E\right) \right|_{{\cal N}=1} . \label{beta-Gtt-E}
\ee
Here the subscript ${\cal N}=1$ indicates that we are just looking at the contributions to $\overline{G}^{(1)}$ and $\delta E$ arising from ${\cal N}=1$ sectors. From \eqref{relat-Gtt} it is obvious that the apparent asymmetry between the three tori in \eqref{beta-Gtt-E} is an artifact of an arbitrary choice and one could have chosen $\overline{G}_{t_2 t_2}^{(1)}$ or $\overline{G}_{t_3 t_3}^{(1)}$ in order to express $\beta$. Note that the right hand side of \eqref{beta-Gtt-E} is actually constant in view of the scaling \eqref{tscalingN1}.

Plugging \eqref{beta-Gtt-E} into \eqref{K_N=1} we obtain for the contribution to the K\"ahler potential from $\mathcal{N}=1$ sectors
\begin{empheq}[box=\fbox]{align}
K^{(1)}_{\mathcal{N}=1} &= \frac{2 \left. \left(4 t_1^2\, \overline{G}_{t_1 t_1}^{(1)}- \delta E\right) \right|_{{\cal N}=1}}{ \sqrt{\tau_0 \tau_1 \tau_2 \tau_3}}   \nonumber \\
&= 2\, e^{2 \Phi_4} \left. \left( 4 t_1^2\, \overline{G}_{t_1 t_1}^{(1)} - \delta E \right) \right|_{{\cal N}=1} \nonumber \\
&= 8\, t_1^2\, \left. G_{t_1 t_1}^{(1)}\right|_{{\cal N}=1}\,.  \label{K_N=1_Final}
\end{empheq}
In the second and the last equality we used \eqref{t0-tau} and \eqref{Einstein_frame_Gtt}, respectively.\footnote{We emphasize that $K^{(1)}$ given in \eqref{K_N=1_Final} is to be understood as a function of the corrected variables $\tau$, even though the second and last line in \eqref{K_N=1_Final} depend on $t_i$ which are related to the tree-level variables $\tau^{(0)}$ via \eqref{t0-tau}-\eqref{t-tau0}, see footnote \ref{interchange}. The same will be the case for the $N=2$ sectors, e.g.\ in \eqref{K_N=2}.} Using \eqref{tau^1_N=1} together with \eqref{alpha_Y_X}, \eqref{relat-Gtt} and \eqref{Y00}-\eqref{Y33}, we obtain the field redefinitions as
\begin{empheq}[box=\fbox]{align}
\left. \tau_0^{(1)}\right|_{\mathcal{N}=1} &= e^{2 \Phi_4}  \Big( \sum_{j=1}^3 t_j^2\, \overline{G}_{t_j t_j}^{(1)} + \sum_{j=1}^3 t_j \overline{G}_{t_0 t_j}^{(1)} - \frac14 \overline{G}_{t_0 t_0}^{(1)}  -\frac12 \delta E  \left. \Big)\right|_{{\cal N}=1} \, \tau_0  \ , \label{tau01N1} \\
\left. \tau_i^{(1)}\right|_{\mathcal{N}=1} &= e^{2 \Phi_4} \Big( \sum_{j=1}^3 t_j^2\, \overline{G}_{t_j t_j}^{(1)} - \sum_{j=1}^3 t_j \overline{G}_{t_0 t_j}^{(1)} + 2 t_i \overline{G}_{t_0 t_i}^{(1)} - \frac14 \overline{G}_{t_0 t_0}^{(1)}  -\frac12 \delta E  \left. \Big) \right|_{{\cal N}=1} \, \tau_i \quad (i \neq 0) \ . \label{tau11N1}
\end{empheq} 
Here we also used \eqref{Einstein_frame_Gtt} in order to express the results in terms of the quantities directly calculable via string amplitudes. Note that the quantities in the brackets of \eqref{tau01N1} and \eqref{tau11N1} are moduli-independent constants for $\mathcal{N}=1$ sectors, cf.\ \eqref{tscalingN1}.  


\subsection{$\mathcal{N}=2$ sectors}  \label{sec.N=2}


\subsubsection{Ansatz for the moduli dependence}
\label{ansatzN2}

In the case of $\mathcal{N}=2$ sectors, besides the normalization of the vertex operators and the loop (dilaton) factor, there are additional moduli dependences (in comparison with \eqref{Gtt_N=1}-\eqref{Gcc_N=1}), coming from Kaluza-Klein (KK) or winding states. We make the following ansatz for the moduli dependence of the $\mathcal{N}=2$ sector contributions to the moduli metric:
\begin{empheq}[box=\fbox]{align}
G_{\tau_i^{(0)} \tau_j^{(0)}}^{(1)}(\tau^{(0)}) &=&   
\sum_{\begin{array}{c} 
{\cal N}=2 \\[-1ex] {\rm sectors} \end{array}} 
 e^{2\Phi_4}\, \frac{\alpha_{ij}^{(m,l)}(U_l) t_l^m}{\tau_i^{(0)} \tau_j^{(0)}} =  
\sum_{\begin{array}{c} 
{\cal N}=2 \\[-1ex] {\rm sectors} \end{array}}  
\frac{\alpha_{ij}^{(m,l)}(U_l) t_l^m}{\tau_i^{(0)} \tau_j^{(0)} \sqrt{\tau_0^{(0)} \tau_1^{(0)} \tau_2^{(0)} \tau_3^{(0)}}}\ , \label{Gtt_N=2} \\
G_{c_i^{(0)} c_j^{(0)}}^{(1)}(\tau^{(0)}) &=& 
\sum_{\begin{array}{c} 
{\cal N}=2 \\[-1ex] {\rm sectors} \end{array}} 
 e^{2\Phi_4}\, \frac{\beta_{ij}^{(m,l)}(U_l) t_l^m}{\tau_i^{(0)} \tau_j^{(0)}} =  
\sum_{\begin{array}{c} 
{\cal N}=2 \\[-1ex] {\rm sectors} \end{array}}  
\frac{\beta_{ij}^{(m,l)}(U_l) t_l^m}{\tau_i^{(0)} \tau_j^{(0)} \sqrt{\tau_0^{(0)} \tau_1^{(0)} \tau_2^{(0)} \tau_3^{(0)}}}\ .  \label{Gcc_N=2}
\end{empheq} 
Obviously $\beta_{ij}$ and $\alpha_{ij}$ are symmetric with respect to $i\leftrightarrow j$. Some remarks concerning the notation are in order. First, each $\mathcal{N}=2$ sector gets contributions from momentum/winding states along a particular torus which is denoted by $l\in {1,2,3}$ in \eqref{Gtt_N=2} and \eqref{Gcc_N=2}. In this way the dependence on the volume $t_l$ and the complex structure $U_l$ of the $l$th torus appears. The functions $\alpha_{ij}^{(m,l)}$ and $\beta_{ij}^{(m,l)}$ result from summing over infinite towers of such KK or winding states and are typically given by (sums of) Eisenstein series. Note that $t_l$ depends on $\tau_i^{(0)}$ via \eqref{t-tau0}. Second, the index $m$ takes on the following values:
\be
m &=& -1 \quad \ ({\rm for\ closed\ string\ winding\ state\ exchange})\ , \nonumber \\
m &=& 1 \qquad ({\rm for\ closed\ string\ KK\ state\ exchange})\ . \label{m1-1}
\ee

The $t$- and $\tau^{(0)}$-scaling in \eqref{Gtt_N=2} can be motivated as follows. Like for the contribution from ${\cal N}=1$ sectors, in string theory one would use the vertex operators for the $t_i$ (with $i \in \{ 0,1,2,3 \}$) and the graviton in order to calculate $\overline G^{(1)}_{t_i t_j}$ and $\delta E$. The ${\cal N}=2$ sector contributions to these quantities can be decomposed as
\be \label{deltaEGbardecompose}
\delta E = \sum_{\begin{array}{c} 
{\cal N}=2 \\[-1ex] {\rm sectors} \end{array}} \delta E^{(m,l)} \qquad , \qquad \overline{G}^{(1)}_{t_i t_j} =  \sum_{\begin{array}{c} 
{\cal N}=2 \\[-1ex] {\rm sectors} \end{array}} 
\overline{G}^{(1)(m,l)}_{t_i t_j}\ . 
\ee
The $t$- and $\tau^{(0)}$-dependence of \eqref{Gtt_N=2} is equivalent to the scaling 
\be \label{tscaling}
\delta E^{(m,l)} \sim t_l^m \ , \quad \overline{G}^{(1)(m,l)}_{t_i t_j} \sim \frac{t^m_l}{t_i t_j}  \ , \quad  \overline{G}^{(1)(m,l)}_{t_0 t_i} \sim \frac{t^m_l}{t_i} \ , \quad \overline{G}^{(1)(m,l)}_{t_0 t_0} \sim t^m_l\ , \quad i,j \in \{ 1,2,3 \}\ .
\ee 
This equivalence could be inferred by interchanging the coordinates $\tau$ and $t$ (cf.\ \eqref{Gtt-Y}-\eqref{X1}) and using \eqref{Einstein_frame_Gtt}. The denominators in \eqref{tscaling} can again be understood from the normalization of the vertex operators, as for the ${\cal N}=1$ sector contributions. Apart from these denominators, the only volume modulus that the quantities $\delta E$ and  $\overline{G}^{(1)(m,l)}_{t_i t_j}$ (for $i \in \{ 0,1,2,3 \}$) can depend on is $t_l$, as this is the only volume modulus on which the KK and winding sums in the string amplitude depend on. Thus, our task is to motivate the scaling of $t_l$ in \eqref{tscaling}. 

For $\delta E^{(m,l)}$ and $\overline{G}^{(1)(m,l)}_{t_0 t_0}$ the $t_l$-scaling follows from the calculation in \cite{Haack:2015pbv}. That paper only considered the 2-point function of gravitons, but concerning the scaling with the volume moduli the calculation for the dilaton would proceed completely analogously, given that the vertex operators only differ by the (volume independent) polarization tensors, cf.\ \eqref{vertexhD}-\eqref{dilatonpolarisation}.

For $\overline{G}^{(1)(m,l)}_{t_i t_j}$ with $i=j=l$ the scaling in \eqref{tscaling} follows from the calculation in \cite{Berg:2005ja}. The scaling of $t_l$ is directly related to the power of the worldsheet parameter in the integrand of the string amplitude (the same holds true for the graviton and dilaton amplitudes). The reason for the fact that the worldsheet parameter only appears with a simple power (besides a simple exponential factor from the KK/winding sum) is that the corrections to the metric from $\mathcal{N}=2$ sectors arise (in the closed string channel) due to the exchange of BPS states. The heavy string oscillators do not contribute and, thus, the integrand of the string amplitude does not contain any theta-functions after spin structure summation, cf.\ (2.64) in \cite{Berg:2005ja}. For the same reason we expect the integrand of the string amplitude (after spin structure summation) to be given by a simple power of the worldsheet parameter for the other cases as well (i.e.\ $i,j \neq l$). Our ansatz \eqref{Gtt_N=2} then amounts to the assumption that this simple power is the same for all $i,j \in \{0,1,2,3 \}$. 

\begin{figure}
\begin{center}
\includegraphics[width=0.375\textwidth]{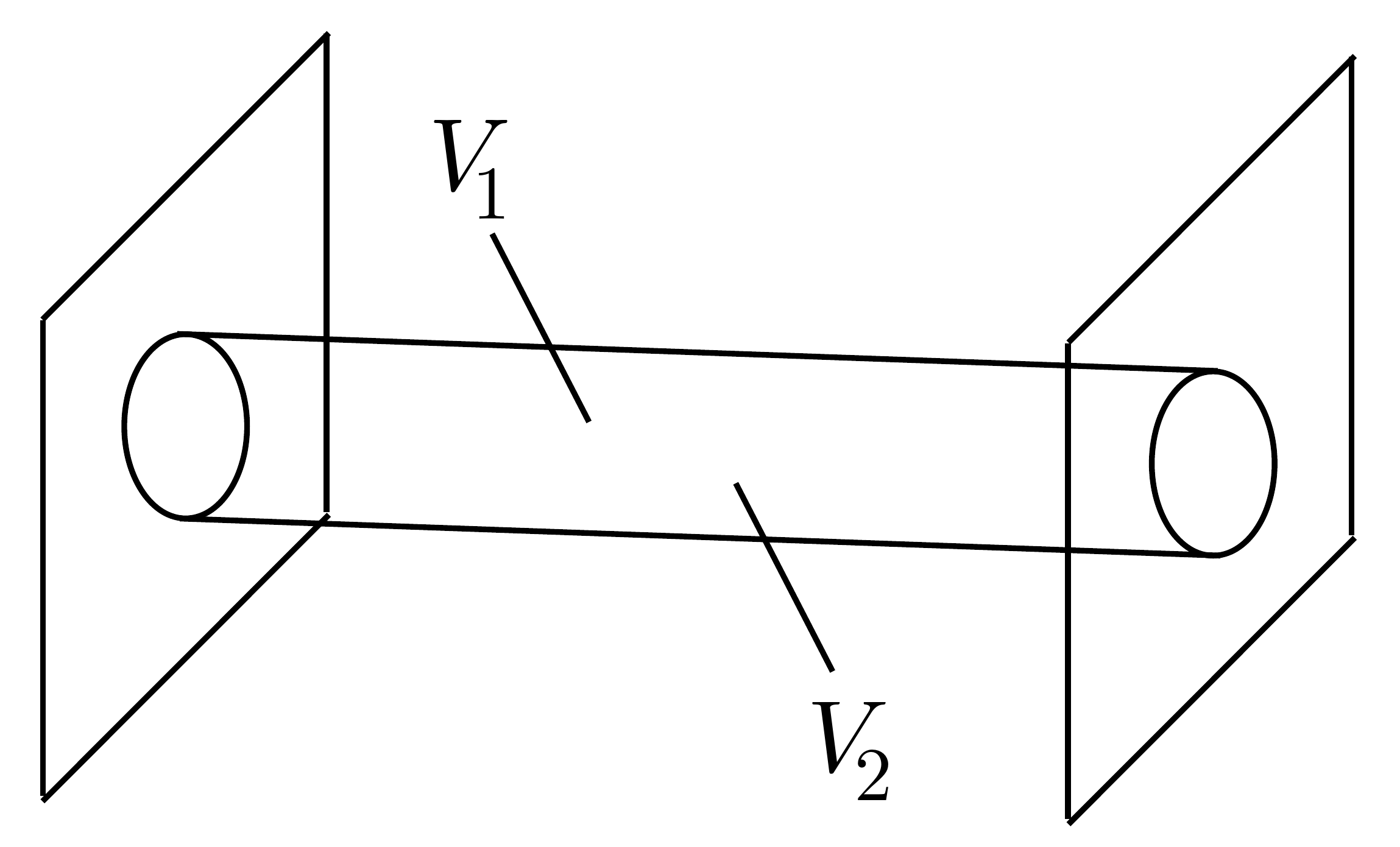}\hspace{0.75cm}
\includegraphics[width=0.5\textwidth]{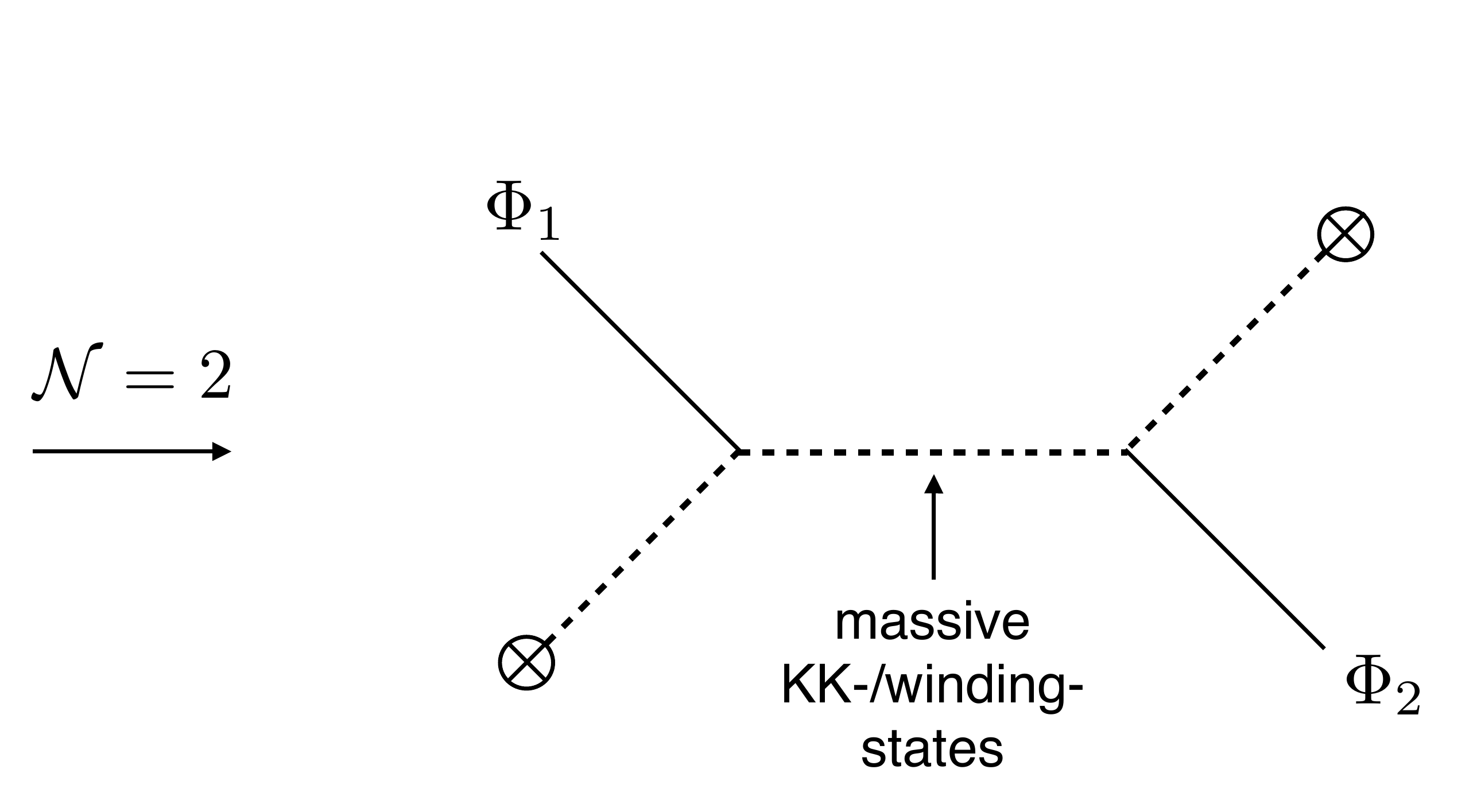}
\caption{Only BPS states contribute in ${\cal N}=2$ sectors. In the string theory picture on the left, $V_1$ and $V_2$ are the vertex operators of any of the moduli and $\Phi_1$ and $\Phi_2$ are the corresponding fields in the field theory picture on the right. Moreover, the crosses on the right hand side stand for the D-brane or O-plane backgrounds.}
\label{KK_winding}
\end{center}
\end{figure}

Alternatively, from a field theory perspective the other cases differ from the case $i=j=l$ only by the 3-point vertices coupling the moduli to the massive BPS-states, cf.\ the right hand side of figure \ref{KK_winding}. 
Thus, our ansatz \eqref{Gtt_N=2} amounts to the assumption that the vertices for the fields $t_i$ with $i \neq l$ have the same $t_l$-dependence as the vertex for $t_l$.\footnote{Two comments are in order here concerning this statement. In the field theory language we interpret the denominators in \eqref{tscaling} (arising from the normalization of the vertex operators in the string theory calculation) as coming from the normalization of the fields and not from the interaction vertex. Moreover, note that some of the interaction vertices might actually be zero.} We leave a verification of this assumption to future work and here content ourselves with the remark that this assumption will allow us to reproduce the complete structure of the K\"ahler potential, found in \cite{Berg:2005ja} for the $\mathbb{Z}_6'$ orientifold using T-duality arguments, cf.\ section \ref{applicationZ6} below. 

Now, assuming \eqref{Gtt_N=2}, the form of \eqref{Gcc_N=2} then follows from the constraints \eqref{cons-cond1} and \eqref{cons-cond3}. This is shown explicitly in appendix \ref{app_beta_ij}.

Based on the above arguments, we have the following sector-decompositions for the 1-loop corrections to the metric components coming from $\mathcal{N}=2$ sectors:
\be
G_{\tau_i^{(0)} \tau_j^{(0)}}^{(1)}(\tau) =
\sum_{\begin{array}{c} 
{\cal N}=2 \\[-1ex] {\rm sectors} \end{array}} 
G_{\tau_i^{(0)} \tau_j^{(0)}}^{(1)|(m,l)}(\tau) \quad , \quad 
G_{c_i^{(0)} c_j^{(0)}}^{(1)}(\tau) =
\sum_{\begin{array}{c} 
{\cal N}=2 \\[-1ex] {\rm sectors} \end{array}} 
G_{c_i^{(0)} c_j^{(0)}}^{(1)|(m,l)}(\tau)
\label{G-decomp_m-l}
\ee
with
\be 
G_{\tau_i^{(0)} \tau_j^{(0)}}^{(1)|(m,l)}(\tau)&=&
 \frac{\alpha_{i j}^{(m,l)}(U_l)\, t_l^m }{\tau_i \tau_j \sqrt{\tau_0 \tau_1 \tau_2 \tau_3}}  
=   \frac{\alpha_{i j}^{(m,l)}(U_l) }{\tau_i \tau_j 
(\tau_0 \tau_l)^{\frac{(1-m)}{2}} } \cdot \left(\frac{\tau_l}{\tau_1 \tau_2 \tau_3}\right)^{\frac{(1+m)}{2}}
\,,  \label{N=2_Gtautau_0} \\
G_{c_i^{(0)} c_j^{(0)}}^{(1)|(m,l)}(\tau)&=&
 \frac{\beta_{i j}^{(m,l)}(U_l)\, t_l^m }{\tau_i \tau_j \sqrt{\tau_0 \tau_1 \tau_2 \tau_3}}  
=   \frac{\beta_{i j}^{(m,l)}(U_l) }{\tau_i \tau_j 
(\tau_0 \tau_l)^{\frac{(1-m)}{2}} } \cdot \left(\frac{\tau_l}{\tau_1 \tau_2 \tau_3}\right)^{\frac{(1+m)}{2}}
\,,  \label{N=2_Gcc_0}
\ee
where we used \eqref{t-tau0}. We make an analogous decomposition of the field redefinitions $\tau_i^{(1)}$ and the 1-loop corrections to the K\"ahler potential, i.e.
\be
\left. \tau_i^{(1)}\right|_{\mathcal{N} = 2}= \sum_{\begin{array}{c} 
{\cal N}=2 \\[-1ex] {\rm sectors} \end{array}}  
\tau_i^{(1)|(m,l)}(\tau) \,,
\ee
\be
K^{(1)}_{\mathcal{N} = 2}(\tau)=\sum_{\begin{array}{c} 
{\cal N}=2 \\[-1ex] {\rm sectors} \end{array}} 
K^{(1)|(m,l)}(\tau) \,. \label{K-sum-(m,l)}
\ee
Then the equations \eqref{eq-ij}-\eqref{eq-ii} and \eqref{K1ij}-\eqref{K1ii} hold for each ${\cal N}=2$ sector (specified by $(m,l)$) separately. 

In the following we would like to follow the strategy again that allowed us to express the field redefinitions and the correction to the K\"ahler potential from ${\cal N}=1$ sectors. Thus, let us pause a moment to recap the steps we took there:
\begin{itemize}
\item Use the consistency conditions \eqref{cons-cond1}-\eqref{cons-cond3} in order to constrain the $\beta$s and $\alpha$s and relate the non-vanishing $\beta$s to the $\alpha$s.
\item Use \eqref{eq-ij}-\eqref{eq-ii} and \eqref{K1ij}-\eqref{K1ii} in order to obtain the field redefinitions and corrections to the K\"ahler potential in terms of the $\alpha$s, employing also the relations between $\beta$s and $\alpha$s found in step 1.
\item Express the result of the second step via $\overline G^{(1)}$ and $\delta E$, using the constraints on the $\alpha$s resulting from the first step.
\end{itemize}


\subsubsection{Constraints on $\alpha$ and $\beta$ from the consistency conditions}
\label{consistency_N2}

Let us introduce indices as 
\be
&& a, b \in \{0,l\} \,,\label{a-b} \\
&& I, J \in \{1,2,3\} \setminus \{l\} \,. \label{I-J}
\ee  
$I$ and $J$ refer to the tori transversal to $l$th torus (i.e.\ to the torus along which a KK/winding sum arises). This separation of the indices obviously depends on the selection of ${\cal N}=2$ sector. 
Now plugging \eqref{N=2_Gtautau_0} and \eqref{N=2_Gcc_0} into the consistency conditions \eqref{cons-cond1}-\eqref{cons-cond3} to get the constraints on $\alpha$ and $\beta$, we obtain (with $a\neq b$ and $I \neq J$):
\begin{itemize}
\item{$m=-1$ sector (closed string channel winding sum):}
\be
\beta_{aa}^{(-1,l)}&=&\alpha_{aa}^{(-1,l)}\ , \label{KK-beta-alpha_aa} \\
\beta_{ab}^{(-1,l)}&=& \frac{1}{2} \left(\alpha_{aa}^{(-1,l)}+\alpha_{bb}^{(-1,l)} \right) - \alpha_{ab}^{(-1,l)}\ , \label{KK-beta-alpha_ab} \\
\beta_{II}^{(-1,l)}&=&2 \alpha_{aI}^{(-1,l)}\ , \label{KK-beta-alpha_II}\\
\beta_{aI} ^{(-1,l)}&=& 0\ , \label{KK-beta-alpha_aI}\\
\beta_{IJ} ^{(-1,l)}&=& 0\ , \label{KK-beta-alpha_IJ} \\ 
\alpha_{aI}^{(-1,l)} &=&\alpha_{bI}^{(-1,l)}\ ,  \label{KK-alpha_aI}\\
\alpha_{II}^{(-1,l)} &=& 0\ , \label{KK-alpha_II} \\
\alpha_{IJ}^{(-1,l)} &=& 0 \label{KK-alpha_IJ} \ .
\ee

\item{$m=1$ sector (closed string channel KK sum):}
\be
\beta_{II}^{(1,l)}&=&\alpha_{II}^{(1,l)}\ , \label{W-beta-alpha_II}\\
\beta_{IJ}^{(1,l)}&=& \frac{1}{2} \left(\alpha_{II}^{(1,l)}+\alpha_{JJ}^{(1,l)} \right) - \alpha_{IJ}^{(1,l)}\ ,  \label{W-beta-alpha_IJ} \\
\beta_{aa}^{(1,l)}&=&2 \alpha_{aI}^{(1,l)}\ , \label{W-beta-alpha_aa} \\
\beta_{aI} ^{(1,l)}&=& 0\ , \label{W-beta-alpha_aI} \\
\beta_{ab} ^{(1,l)}&=& 0\ ,  \label{W-beta-alpha_ab} \\ 
\alpha_{aI}^{(1,l)} &=&\alpha_{aJ}^{(1,l)}\ , \label{W-alpha_aI}\\
\alpha_{aa}^{(1,l)} &=& 0\ ,  \label{W-alpha_aa}\\
\alpha_{ab}^{(1,l)} &=& 0 \label{W-alpha_ab}\ .
\ee
Note that the $m=1$ conditions can be obtained from the $m=-1$ conditions by replacing $a\leftrightarrow I$ and $b \leftrightarrow J$.   
\end{itemize}


\subsubsection{Field redefinitions and K\"ahler potential} 
\label{fieldredef_K_N=2}

We now solve the equations for the field redefinitions and the correction to the K\"ahler potential, \eqref{eq-ij}-\eqref{eq-ii} and \eqref{K1ij}-\eqref{K1ii}, using the relations \eqref{KK-beta-alpha_aa}-\eqref{W-alpha_ab} wherever applicable. We relegate the calculational details to appendix \ref{detailsN2}.

For the field redefinitions we obtain
\begin{empheq}[box=\fbox]{align}
\tau_a^{(1)|(-1,l)}(\tau) &= \frac{2\left(\alpha_{bb}^{(-1,l)} - 2 \alpha_{ab}^{(-1,l)}\right)}{\tau_b} \,,\label{tau_a^(-1,l)}  \\
\tau_I^{(1)|(-1,l)}(\tau) &=   
\frac{2 \tau_I \left(\alpha_{II}^{(-1,l)} -2 \alpha_{aI}^{(-1,l)}\right) }{
 \tau_0 \tau_l } =  \frac{- 4 \tau_I \, \alpha_{aI}^{(-1,l)} }{
 \tau_0 \tau_l } \,, \\
\tau_a^{(1)|(1,l)}(\tau) &= 
\frac{2 \tau_a  \left(\alpha_{aa}^{(1,l)} -2 \alpha_{aI}^{(1,l)}\right) }{
 \tau_I \tau_J  } =  \frac{- 4 \tau_a \, \alpha_{aI}^{(1,l)} }{
 \tau_I \tau_J }\,,\\
\tau_I^{(1)|(1,l)}(\tau) &= \frac{2\left(\alpha_{JJ}^{(1,l)} - 2 \alpha_{IJ}^{(1,l)}\right)}{\tau_J} \, \label{tau_I^(1,l)}
\end{empheq}
with $a \neq b$ and $I \neq J$.

Let us turn to \eqref{K1ij}-\eqref{K1ii}, which give for each sector $(m,l)$
\be
\frac{1}{4}\,\frac{\partial^2 K^{(1)|(m,l)}(\tau)}{\partial \tau_i \partial \tau_j} &=& 
     G_{c_i^{(0)} c_j^{(0)}}^{(1)|(m,l)}(\tau)       \qquad \qquad \qquad \text{if } i\neq j  \,, \label{K1_N=2}  \\ 
\frac{1}{4}\,   \frac{\partial^2 K^{(1)|(m,l)}(\tau)}{\partial \tau_i \partial \tau_i} &=&  G_{c_i^{(0)} c_i^{(0)}}^{(1)|(m,l)}(\tau) + 
\frac{\tau_i^{(1)|(m,l)}(\tau)}{2 \tau_i^3}   
\label{K2_N=2}\,. 
\ee
These equations are solved by (cf.\ appendix \ref{detailsN2} for more details)
\begin{empheq}[box=\fbox]{align}
K^{(1)|(-1,l)}(\tau)&= \frac{2 \left(\alpha_{aa}^{(-1,l)}+\alpha_{bb}^{(-1,l)} - 2 \alpha_{ab}^{(-1,l)}\right)}{ \tau_a \tau_b} 
= \frac{4 \beta_{ab}^{(-1,l)}}{ \tau_a \tau_b}\ ,\label{K_N=2_KK} \\
K^{(1)|(1,l)}(\tau)&= \frac{2 \left(\alpha_{II}^{(1,l)}+\alpha_{JJ}^{(1,l)} - 2 \alpha_{IJ}^{(1,l)}\right)}{ \tau_I \tau_J}
= \frac{4 \beta_{IJ}^{(1,l)}}{ \tau_I \tau_J} \ . \label{K_N=2_W}
\end{empheq}
Recall our notation for the indices: $a\neq b \in \{0,l\}$ and $I\neq J \in \{1,2,3\} \setminus \{l\}$. The total ${\cal N}=2$-contribution to the K\"ahler potential is the sum of all the ${\cal N}=2$-sectors as given in \eqref{K-sum-(m,l)}.

From the consistency conditions, we have found constraints on $\alpha_{ij}^{(m,l)}$, \eqref{KK-alpha_aI}-\eqref{KK-alpha_IJ} and \eqref{W-alpha_aI}-\eqref{W-alpha_ab}. Here we will investigate what these constraints imply for the metric components and use our findings in order to express the field redefinitions and the corrections to the K\"ahler potential in terms of $\overline{G}$ and $\delta E$.

From \eqref{Gtt-Y} and \eqref{Gtt_N=2}, we have 
\be
\alpha_{ij}^{(m,l)} = e^{-2 t_0} \left( t_l^{-m} \right) Y_{ij}^{(1)|(m,l)}\ ,  \label{alpha-Y}
\ee    
where $Y_{ij}^{(1)|(m,l)}$ is the $(m,l)$-sector component of $Y_{ij}^{(1)}$, i.e.
\be
Y_{ij}^{(1)}= \sum_{\begin{array}{c} 
{\cal N}=2 \\[-1ex] {\rm sectors} \end{array}} 
Y_{ij}^{(1)|(m,l)}\,.
\ee

Now we solve the constraint equations \eqref{KK-alpha_aI}-\eqref{KK-alpha_IJ} and \eqref{W-alpha_aI}-\eqref{W-alpha_ab}, using \eqref{Y00}-\eqref{Y33}.
For the $(m,l)$ sector we find 
\bea
t_l^2 \, G_{t_l t_l}^{(1)|{(m,l)}} &=& - \frac{G_{t_0 t_0}^{(1)|{(m,l)}}}{4} + m\, t_l \,G_{t_0 t_l}^{(1)|{(m,l)}}\ , \label{Gtll_N=2} \\
t_I^2 \, G_{t_I t_I}^{(1)|{(m,l)}} &=& t_J^2 \, G_{t_J t_J}^{(1)|{(m,l)}} = -m\, t_I t_J G_{t_I t_J}^{(1)|{(m,l)}} \ ,   \label{Gtii_N=2} \\
m \,G_{t_0 t_I}^{(1)|{(m,l)}} &=& 2 t_l G_{t_l t_I}^{(1)|{(m,l)}}\ . \label{Gfti_N=2} 
\eea
Using \eqref{Einstein_frame_Gtt}, the relations \eqref{Gtll_N=2}-\eqref{Gfti_N=2} for the Einstein-frame metrics can be expressed in terms of $\overline G$ and $\delta E$, i.e.\ the quantities which are directly calculable via string 2-point amplitudes. This results in
\bea
t_l^2 \, \overline{G}_{t_l t_l}^{(1)|{(m,l)}} &=& - \frac{\overline G_{t_0 t_0}^{(1)|{(m,l)}}}{4} +m\, t_l \overline{G}_{t_0 t_l}^{(1)|(m,l)} + \frac{\delta E^{(m,l)}}{2} \ , \label{tilde_G_tt_ll} \\
t_I^2 \, \overline{G}_{t_I t_I}^{(1)|{(m,l)}} &=& t_J^2 \, \overline{G}_{t_J t_J}^{(1)|{(m,l)}} = \frac{\delta E^{(m,l)}}{4}  -m\, t_I t_J \overline{G}_{t_I t_J}^{(1)|{(m,l)}}   
\qquad \text{for} \quad  I \neq J \ , \label{tilde_G_tt_ii} \\
\overline{G}_{t_0 t_I}^{(1)|(m,l)} &=& 2 m\,  t_l\, \overline{G}_{t_l t_I}^{(1)|{(m,l)}}    \label{dE/dt_i} \ .
\eea

Plugging \eqref{alpha-Y} into \eqref{KK-beta-alpha_ab} and \eqref{W-beta-alpha_IJ}, and using \eqref{Y0-1}-\eqref{Y1-2} with \eqref{Gtll_N=2}-\eqref{Gfti_N=2},  
we have
\be
\beta_{0l}^{(-1,l)} &= 2\, e^{-2 t_0} t_l t_I^2\, G_{t_I t_I}^{(1)|{(-1,l)}} &= 2\, t_1 t_2 t_3\, \overline{G}_{t_I t_J}^{(1)|{(-1,l)}} \,,\\
\beta_{IJ}^{(1,l)} &=  2\, e^{-2 t_0} t_l^{-1} t_I^2\, G_{t_I t_I}^{(1)|{(1,l)}} &= - 2\, t_l^{-1} t_I t_J\, \overline{G}_{t_I t_J}^{(1)|{(1,l)}} \,.
\ee 
Thus, from \eqref{K_N=2_KK}-\eqref{K_N=2_W} we obtain the contributions to the K\"ahler potential,
\be
K^{(1)|(- 1,l)}(\tau) &=& \frac{8\,  t_1 t_2 t_3\, \overline{G}_{t_I t_J}^{(1)|{(-1,l)}}}{\tau_0 \tau_l}= 
8 \,e^{2 \Phi_4} t_I t_J  \overline{G}_{t_I t_J}^{(1)|(-1,l)}   \label{K_N=2_KK_0} \,, \\
K^{(1)|( 1,l)}(\tau) &=& -\, \frac{8 \,t_I t_J\, \overline{G}_{t_I t_J}^{(1)|(1,l)}}{t_l \tau_I \tau_J} =
- \, 8\, e^{2 \Phi_4} t_I t_J\overline{G}_{t_I t_J}^{(1)|{(1,l)}}    \label{K_N=2_W_0} \,
\ee
with $I, J\in\{1,2,3\} \setminus \{l\} $ and $I\neq J$. The second equalities follow from \eqref{t0-tau} and \eqref{t-tau0}. 

It may be more useful to express the K\"ahler potential in terms of diagonal metric components, rather than off-diagonal components. That is, \eqref{K_N=2_KK_0} and \eqref{K_N=2_W_0} can be written, using \eqref{tilde_G_tt_ii}, as 
\be
K^{(1)|(-1,l)}(\tau) &=& \frac{ 2 t_l \left(4 t_I^2  \overline{G}_{t_I t_I}^{(1)|(-1,l)}- \delta E^{(-1,l)} \right)}{ \tau_0 \tau_l}  \,, \\
K^{(1)|(1,l)}(\tau) &=&  \frac{ 2 t_l^{-1} \left(4 t_I^2  \overline{G}_{t_I t_I}^{(1)|(1,l)}- \delta E^{(1,l)} \right)}{ \tau_I \tau_J} 
\ee
with $I, J\in\{1,2,3\} \setminus \{l\} $ and $I\neq J$. Note that the numerators above are independent of $t_i$ as can be inferred using \eqref{tscaling}. More compactly the above can be written as 
\begin{empheq}[box=\fbox]{align}
K^{(1)|(m,l)}(\tau) & = 
2 \,e^{2 \Phi_4} \left(4 t_I^2  \overline{G}_{t_I t_I}^{(1)|(m,l)}- \delta E^{(m,l)} \right)
=  8 \,  t_I^2 \, G_{t_I t_I}^{(1)|(m,l)}  \label{K_N=2} 
\end{empheq}
with any $I\in\{1,2,3\} \setminus \{l\} $. In the second equality we used \eqref{Einstein_frame_Gtt}. Note that the above is of the same form as the contribution from $\mathcal{N}=1$ sectors, cf.\ \eqref{K_N=1_Final}.\footnote{ \eqref{K_N=1_Final} could have been written in terms of any one of the 3 tori, rather than in terms of the first one. \eqref{K_N=2} in the case of $\mathcal{N}=2$ has the same form as \eqref{K_N=1_Final}, but is restricted  to a direction orthogonal to the KK/winding direction.} Actually, any 4-dimensional toroidal $\mathbb{Z}_N$ orientifold with minimal supersymmetry has one torus which is orthogonal to the KK/winding direction in every ${\cal N}=2$ orbifold sector. In the notation of table 2 in \cite{Aldazabal:1998mr}, this is the first torus. In that case it follows from \eqref{K_N=1_Final} and \eqref{K_N=2} that the complete 1-loop correction to the K\"ahler potential is determined by the combination $4 t_1^2 \overline{G}_{t_1 t_1}^{(1)} -  \delta E$, i.e.  
\begin{empheq}[box=\fbox]{align}
K^{(1)}= 2 e^{2 \Phi_4} \left(4 t_1^2 \overline{G}_{t_1 t_1}^{(1)} -  \delta E \right) =  8 t_1^2 G_{t_1 t_1}^{(1)}  \qquad \quad \text{for $\mathbb{Z}_N$ models.}  \label{Ksimple}
\end{empheq}


\subsection{An observation on the structure of the field redefinitions and the corrections to the K\"ahler potential}
\label{observation}

Before applying our results to a concrete example in the following section, we would like to make a general comment about the structure of the field redefinitions and the corrections to the K\"ahler potential. The tree-level K\"ahler potential \eqref{K0T0} can be expressed in terms of the corrected variables $\tau$ according to 
\be
K^{(0)} (T^{(0)}, \bar T^{(0)})&=& - \sum_{i=0}^3\ln \left(T_i^{(0)} - \bar T_i^{(0)}\right)  
= -\ln\left[16\, \tau_0^{(0)} \tau_1^{(0)} \tau_2^{(0)} \tau_3^{(0)}\right] = -\ln\left[16\, \prod_{i=0}^3 \left(\tau_i-\tau_i^{(1)}\right)\right]
\nonumber \\
&=&  K^{(0)} (T, \bar T) + \frac{\tau_0^{(1)}}{\tau_0}  +\frac{\tau_1^{(1)}}{\tau_1} + \frac{\tau_2^{(1)}}{\tau_2} + \frac{\tau_3^{(1)}}{\tau_3} + {\rm higher\ orders}  \ .
\ee
Comparing this with \eqref{fullK}, we see that {\it if} it turns out that 
\be \label{ABCDconds-0}
\sum_{i=0}^3 \frac{\tau_i^{(1)}}{\tau_i} = K^{(1)}(\tau) 
\ee
{\it then} the 1-loop correction to the K\"ahler potential could be interpreted as being generated solely from expressing the tree-level K\"ahler potential \eqref{K0T0} in terms of the corrected K\"ahler variables, as was assumed in the analysis of \cite{Conlon:2010ji}. Whether this really happens depends on the explicit form of the field redefinitions and the corrections to the K\"ahler potential (including the exact coefficients to be determined by string theory), but it is already interesting to notice that the field redefinitions have the right {\it structure} for this to have a chance to work out. This can be seen from \eqref{tau^1_N=1} and \eqref{K_N=1} for the $\mathcal{N}=1$ sectors and \eqref{tau_a^(-1,l)}-\eqref{tau_I^(1,l)} 
and \eqref{K_N=2_KK}-\eqref{K_N=2_W} for the $\mathcal{N}=2$ sectors. Using these equations (and the constraints \eqref{alphas-N=1}) the conditions \eqref{ABCDconds-0} can be expressed in terms of $\alpha$s, resulting in
\be
&&\alpha_{01} + \alpha_{12} +  \alpha_{23} + \alpha_{30} = 0\ ,  \label{Con-Ped-N=1} \\
&&\alpha_{01}^{(-1,l)}+\alpha_{02}^{(-1,l)}+\alpha_{03}^{(-1,l)}=0\ , \label{Con-Ped-W} \\
&& \alpha_{I J}^{(1,l)}+\alpha_{0 I}^{(1,l)}+\alpha_{l I}^{(1,l)}=0\ , \qquad \text{with} \quad I \neq J \in \{1,2,3\} \setminus  l \ . \label{Con-Ped-KK}
\ee
Upon using \eqref{alpha_Y_X}, \eqref{alpha-Y}, \eqref{Y01}-\eqref{Y03}, \eqref{Y12}-\eqref{Y13}, \eqref{Y23} and \eqref{Einstein_frame_Gtt} and imposing the constraints \eqref{relat-tilde-Gtt} and \eqref{tilde_G_tt_ll}-\eqref{dE/dt_i},
the conditions \eqref{Con-Ped-N=1}-\eqref{Con-Ped-KK} can be solved by
\be \label{condtauKN1}
4 t_1^2 \overline{G}_{t_1 t_1}^{(1)}\Big|_{\mathcal{N}=1} = \overline{G}_{t_0 t_0}^{(1)} \Big|_{\mathcal{N}=1}
\ee   
for the $\mathcal{N}=1$ sectors and 
\be \label{condtauKN2}
4 t_I^2 \overline{G}_{t_I t_I}^{(1)|(m,l)} = \overline{G}_{t_0 t_0}^{(1)|(m,l)}\, \qquad \text{with} \quad I  \in \{1,2,3\} \setminus  l \,
\ee
for  the $\mathcal{N}=2$ sectors. Thus, the assumption of \cite{Conlon:2010ji} mentioned above is equivalent to the conditions \eqref{condtauKN1} and \eqref{condtauKN2}, which are verifiable by direct string calculations. As mentioned below \eqref{K_N=2}, in the case of a $\mathbb{Z}_N$ orientifold the first torus never supports KK or winding states in ${\cal N}=2$ sectors and, thus, in that case one can summarize the two conditions \eqref{condtauKN1} and \eqref{condtauKN2} by the very simple condition\footnote{Note that condition \eqref{4G11=G00} is meant to hold for each sector, i.e.\ $4 t_1^2 \overline{G}_{t_1 t_1}^{(1)}\Big|_{\mathcal{N}=1} = \overline{G}_{t_0 t_0}^{(1)} \Big|_{\mathcal{N}=1}$ for the $\mathcal{N}=1$ sectors and  $4 t_1^2 \overline{G}_{t_1 t_1}^{(1)|(m,l)} = \overline{G}_{t_0 t_0}^{(1)|(m,l)} $ for the $\mathcal{N}=2$ sectors.} 
\be 
4 t_1^2 \overline{G}_{t_1 t_1}^{(1)} = \overline{G}_{t_0 t_0}^{(1)} \qquad \quad \text{for $\mathbb{Z}_N$ models}\ . \label{4G11=G00}
\ee


\section{Application: $\mathbb{Z}_6'$ orientifold}
\label{applicationZ6}

Let us take the example of the $\mathbb{Z}_6'$ orientifold with twist vector $v=(\frac{1}{6}, -\frac{1}{2}, \frac{1}{3})$ and work out the explicit form of the field redefinitions and K\"ahler potential, using the results obtained above. The $\mathbb{Z}_6'$ orientifold has both ${\cal N}=1$ and ${\cal N}=2$ sectors. The moduli dependence of the field redefinitions and the correction to the K\"ahler potential from ${\cal N}=1$ sectors was given above in equations \eqref{tau^1_N=1} and \eqref{K_N=1} (with constant $\alpha$s and $\beta$s). These results can also be expressed in terms of quantities calculable via string amplitudes, cf.\ \eqref{K_N=1_Final}-\eqref{tau11N1}. For the contributions from ${\cal N}=2$ sectors we can read off the moduli dependence from \eqref{tau_a^(-1,l)}-\eqref{tau_I^(1,l)} and \eqref{K_N=2_KK}-\eqref{K_N=2_W} for the field redefinitions and the correction to the K\"ahler potential, respectively. For the $\mathbb{Z}_6'$ orientifold, the ${\cal N}=2$ sectors are $(m,l) =\{(1,2),(-1,2),(-1,3) \}$ (cf.\ table 3 in \cite{Haack:2015pbv}), which means $\alpha^{(\pm1,1)}=0$ and $\alpha^{(1,3)}=0$, so that the field redefinitions and the correction to the K\"ahler potential read\be
\tau_0^{(1)}\Big|_{\mathcal{N}=2}&=& \frac{2\left(\alpha_{22}^{(-1,2)} - 2 \alpha_{02}^{(-1,2)}\right)}{\tau_2} +
\frac{2\left(\alpha_{33}^{(-1,3)} - 2 \alpha_{03}^{(-1,3)}\right)}{\tau_3}-
 \frac{4 \tau_0 \, \alpha_{03}^{(1,2)} }{\tau_1 \tau_3 }
 \,, \label{Z6-tau-0} \\
\tau_1^{(1)}\Big|_{\mathcal{N}=2}&=& - \frac{ 4 \tau_1 \, \alpha_{01}^{(-1,2)} }{\tau_0 \tau_2 } -\frac{ 4 \tau_1 \, \alpha_{01}^{(-1,3)} }{\tau_0 \tau_3 } 
+ \frac{2\left(\alpha_{33}^{(1,2)} - 2 \alpha_{13}^{(1,2)}\right)}{\tau_3} \,,\label{Z6-tau-1} \\
\tau_2^{(1)}\Big|_{\mathcal{N}=2}&=& \frac{2\left(\alpha_{00}^{(-1,2)} - 2 \alpha_{02}^{(-1,2)}\right)}{\tau_0} 
 - \frac{4 \tau_2 \, \alpha_{02}^{(-1,3)} }{\tau_0 \tau_3 } - \frac{ 4 \tau_2 \, \alpha_{23}^{(1,2)} }{\tau_1 \tau_3 }\,,\label{Z6-tau-2}\\
\tau_3^{(1)}\Big|_{\mathcal{N}=2}&=& -\frac{4 \tau_3 \, \alpha_{03}^{(-1,2)} }{\tau_0 \tau_2 }+
 \frac{2\left(\alpha_{00}^{(-1,3)} - 2 \alpha_{03}^{(-1,3)}\right)}{\tau_0} +
 \frac{2\left(\alpha_{11}^{(1,2)} - 2 \alpha_{13}^{(1,2)}\right)}{\tau_1}  \label{Z6-tau-3}
\ee  
and 
\be
K^{(1)}_{\mathcal{N}=2} = \frac{2 \left(\alpha_{00}^{(-1,2)}+\alpha_{22}^{(-1,2)} - 2 \alpha_{02}^{(-1,2)}\right)}{ \tau_0 \tau_2}
+ \frac{2 \left(\alpha_{11}^{(1,2)}+\alpha_{33}^{(1,2)} - 2 \alpha_{13}^{(1,2)}\right)}{ \tau_1 \tau_3}
+ \frac{2 \left(\alpha_{00}^{(-1,3)}+\alpha_{33}^{(-1,3)} - 2 \alpha_{03}^{(-1,3)}\right)}{ \tau_0 \tau_3} \,. \label{K-Z6} \nonumber \\
\ee 
Again these results can also be expressed in terms of quantities calculable via string amplitudes. Using \eqref{K_N=2} for the K\"ahler potential results in
\be \label{KN2Z6}
K_{\mathcal{N}=2}^{(1)}= 8 t_1^2 \, \left[  G_{t_1 t_1}^{(1)|(-1,2)}  + G_{t_1 t_1}^{(1)|(1,2)} + G_{t_1 t_1}^{(1)|(-1,3)}\right] 
= 2e^{2 \Phi_4} \left. \left( 4 t_1^2 \overline{G}_{t_1 t_1}^{(1)} -  \delta E  \right)\right|_{\mathcal{N}=2} \ .
\ee 
The field redefinitions can also be expressed explicitly in terms of $\overline{G}$ and $\delta E$. The result is rather lengthy and we give the details in appendix \ref{redef_Z6'}.

Let us finally collect all the results for the $\mathbb{Z}_6'$ orientifold, including both ${\cal N}=1$ and ${\cal N}=2$ sectors, and using the relation 
\be \label{Eisenstein}
\alpha_{ij}^{(m,l)} \sim E_2 (-m U_l^{-m}) \qquad , \qquad E_2(U) \equiv \sum_{(m,n) \neq (0,0)} \frac{({\rm Im}(U))^2}{|m + n U|^4}
\ee
between the $\alpha$s and the non-holomorphic Eisenstein series $E_2$, which is known from explicit string calculations, cf.\ \cite{Berg:2005ja,Haack:2015pbv} for instance.\footnote{In writing down \eqref{Eisenstein} we assumed for simplicity that all the D5-branes are sitting at the origin of the second torus. Otherwise the functions $\alpha$ (and $\beta$) would be more complicated, involving also the distances between different D5-branes along the second torus. If all the D5-branes are sitting at the origin of the second torus the tadpoles are not cancelled locally and we neglected backreaction effects in \eqref{Eisenstein}. Of course, the formulas \eqref{Z6-tau-0}-\eqref{K-Z6} are equally valid for other D5-brane configurations, just that the $\alpha_{ij}^{(m,2)}$ would involve sums of Eisenstein series similar to formula (D.35) in \cite{Haack:2008yb} (which is in the T-dual frame though and includes also the T-dual of Wilson-lines on the D9-branes).} For the field redefinitions (expressed in terms of uncorrected fields $\tau^{(0)}$, cf.\ footnote \ref{interchange}) we have
\begin{empheq}[box=\fbox]{align}
\tau_0^{(1)} & =  a_1 \sqrt{\frac{\tau_0^{(0)}}{\tau_1^{(0)} \tau_2^{(0)} \tau_3^{(0)}}} + a_2 \frac{E_2(U_2)}{\tau_2^{(0)}} + a_3 \frac{\tau_0^{(0)} E_2(-1/U_2)}{\tau_1^{(0)} \tau_3^{(0)}} + a_4 \frac{E_2(U_3)}{\tau_3^{(0)}} \ ,\label{deltatau0}\\
\tau_1^{(1)} & =  a_5 \sqrt{\frac{\tau_1^{(0)}}{\tau_0^{(0)} \tau_2^{(0)} \tau_3^{(0)}}} + a_6 \frac{\tau_1^{(0)} E_2(U_2)}{\tau_0^{(0)} \tau_2^{(0)}} + a_7 \frac{E_2(-1/U_2)}{\tau_3^{(0)}} + a_8 \frac{\tau_1^{(0)} E_2(U_3)}{\tau_0^{(0)} \tau_3^{(0)}}\ , \label{deltatau1}\\
\tau_2^{(1)} & =  a_9 \sqrt{\frac{\tau_2^{(0)}}{\tau_0^{(0)} \tau_1^{(0)} \tau_3^{(0)}}} + a_{10} \frac{E_2(U_2)}{\tau_0^{(0)}} + a_{11} \frac{\tau_2^{(0)} E_2(-1/U_2)}{\tau_1^{(0)} \tau_3^{(0)}} + a_{12} \frac{\tau_2^{(0)} E_2(U_3)}{\tau_0^{(0)} \tau_3^{(0)}}\ , \label{deltatau2}\\
\tau_3^{(1)} & =  a_{13} \sqrt{\frac{\tau_3^{(0)}}{\tau_0^{(0)} \tau_1^{(0)} \tau_2^{(0)}}} + a_{14} \frac{\tau_3^{(0)} E_2(U_2)}{\tau_0^{(0)} \tau_2^{(0)}} + a_{15} \frac{E_2(-1/U_2)}{\tau_1^{(0)}} + a_{16} \frac{E_2(U_3)}{\tau_0^{(0)}} \label{deltatau3}
\end{empheq}
and for the K\"ahler potential of the dilaton and the untwisted metric moduli (including also the complex structure $U_2$ of the second torus, which is also a modulus field in the low-energy effective action) 
\begin{empheq}[box=\fbox]{align} \label{Kconjecture}
K = & -\ln (T_0 - \bar T_0) - \ln[(T_1 - \bar T_1) (T_2 - \bar T_2) (T_3 - \bar T_3)] - \ln[(U_2 - \bar U_2)]  \nonumber \\
& - b_1 \chi \zeta(3) \sqrt{\frac{(T_1-\bar{T}_1)(T_2-\bar{T}_2)(T_3-\bar{T}_3)}{(T_0-\bar T_0)^3}} \nonumber \\
& + b_2 \frac{E_2(U_2)}{(T_0-\bar T_0) (T_2-\bar{T}_2)} + b_3 \frac{E_2(-1/U_2)}{(T_1-\bar{T}_1) (T_3-\bar{T}_3)} + b_4 \frac{E_2(U_3)}{(T_0-\bar T_0) (T_3-\bar{T}_3)} \nonumber \\
& + b_5 \frac{1}{\sqrt{(T_0-\bar T_0)(T_1-\bar{T}_1)(T_2-\bar{T}_2)(T_3-\bar{T}_3)}}\ ,
\end{empheq}
where we also included the $\alpha'$-correction from sphere level \cite{Becker:2002nn} in the second row (the $b_1$-term). All the coefficients $a_i$ in \eqref{deltatau0}-\eqref{deltatau3} and $b_i$ in \eqref{Kconjecture} are constants that have to be determined by comparing to concrete string theory calculations (and some of them might actually turn out to be zero). For the $a_i$ this can be done by employing \eqref{tau01N1}-\eqref{tau11N1} and \eqref{Z6-tau-0}-\eqref{Z6-tau-3} together with the formulas of appendix \ref{redef_Z6'}. The coefficients $b_2, \ldots , b_5$ can be obtained using \eqref{K_N=1_Final} and \eqref{KN2Z6}. Note that determining the 1-loop corrections to the K\"ahler potential $K$ is in general much simpler than determining the 1-loop field redefinitions and only requires to calculate the combination $4 t_1^2 \overline{G}_{t_1 t_1}^{(1)} - \delta E$, cf.\ \eqref{Ksimple}. For the field redefinitions, the first terms of \eqref{deltatau0}-\eqref{deltatau3} each arise from the ${\cal N}=1$ sectors of ${\cal A, M, K}$ and ${\cal T}$, whereas the further terms arise from ${\cal N}=2$ sectors of ${\cal A, M}$ and ${\cal K}$. The last term of \eqref{deltatau0} (i.e.\ the one proportional to $a_4$) is the analog of the field redefinition discussed by \cite{Antoniadis:1996vw} in the context of a $\mathbb{T}^2 \times {\rm K}3$-compactification (cf.\ appendix \ref{sec:t2k3}). In the K\"ahler potential (\ref{Kconjecture}), the terms in the third row arise from the ${\cal N}=2$ sectors of ${\cal A, M}$ and ${\cal K}$ and the term in the last row has its origin in the ${\cal N}=1$ sectors of ${\cal A, M, K}$ and ${\cal T}$.

 
\section{Summary and outlook}
\label{summary}

In this paper we have considered the field redefinitions and the K\"ahler potential at string 1-loop for a particular class of string theory models (4-dimensional toroidal type IIB orientifolds with ${\cal N}=1$ supersymmetry) and for a particular subsector of fields (the 4-dimensional dilaton and the diagonal untwisted K\"ahler moduli, i.e.\ the K\"ahler moduli related to the volumes of the three 2-tori). 

The redefinitions of the field variables are required by supersymmetry, in order to make the K\"ahler structure of the scalar metric manifest at 1-loop order. In addition to supersymmetry we made use of perturbative axionic shift symmetries and a particular ansatz for the form of the 1-loop corrections to the metric which is suggested by concrete string calculations. These constraints allowed us to obtain the general structure of the field redefinitions and simultaneously of the K\"ahler potential at 1-loop order. The explicit form of the field redefinitions and the K\"ahler potential (i.e.\ with the exact coefficients) would now, in a second step, require more concrete input from string theory, via string scattering amplitudes. 

The most important results concerning the general structure of the 1-loop field redefinitions can be found in equations \eqref{tau^1_N=1} and \eqref{tau_a^(-1,l)}-\eqref{tau_I^(1,l)} for the contributions from ${\cal N}=1$ and ${\cal N}=2$ sectors, respectively. The notation for the ${\cal N}=2$ sectors is explained below \eqref{Gcc_N=2} and at the beginning of section \ref{consistency_N2}. Moreover, the $\alpha_{ij}$ of the ${\cal N}=1$ sectors are constants to be determined by string theory and the $\alpha_{ij}^{(m,l)}$ of the ${\cal N}=2$ sectors are functions of the complex structure and are given by  (sums of) Eisenstein series with coefficients again to be determined by string theory. Concerning the general structure of the K\"ahler potential, our results are given in \eqref{K_N=1} and \eqref{K_N=2_KK}-\eqref{K_N=2_W}, for the contributions from ${\cal N}=1$ and ${\cal N}=2$ sectors, respectively. We then applied these formulas to the $\mathbb{Z}_6'$-orientifold in section \ref{applicationZ6}. We would like to emphasize, though,  that our general results about the 1-loop field redefinitions and the corrections to the K\"ahler potential hold for an arbitrary 4-dimensional toroidal type IIB orientifold with ${\cal N}=1$ supersymmetry.

The general structure of the K\"ahler potential in the case of the $\mathbb{Z}_6'$-orientifold is not new and was already given in \cite{Berg:2005ja}. Here we derived it in a very different way, confirming the moduli dependence inferred in \cite{Berg:2005ja} via T-duality arguments. Our method has several advantages. First, it is very general, applicable to any 4-dimensional toroidal type IIB orientifold with minimal supersymmetry and it can also be easily generalized to the case of ${\cal N}=2$ supersymmetry, cf.\ appendix \ref{sec:t2k3}. We always used the language of an orientifold with D9/D5-branes, but the results for the 1-loop field redefinitions and the corrections to the K\"ahler potential can be interpreted for the case of D3/D7-branes as well, cf.\ the comments at the end of section \ref{effectiveaction}.\footnote{It was argued in \cite{Antoniadis:2018hqy} that in models with D7-branes there can be additional 1-loop contributions to the K\"ahler potential (not covered by our metric ansatz), albeit for compactifications that do not correspond to free conformal field theories (thus, excluding the toroidal orientifolds discussed in our paper), that have localized string tree-level corrections to the Einstein-Hilbert term and that have local tadpoles.} Second, our method allows us to obtain also the general structure of the field redefinitions. Third, the 1-loop field redefinitions and corrections to the K\"ahler potential can straightforwardly be expressed in terms of quantities that are directly calculable via string scattering amplitudes (i.e.\ $\delta E$ and $\overline{G}^{(1)}_{t_i t_j}$ appearing in the effective action \eqref{1-loop-EF-action-string-1}). We consider these expressions another important outcome of our analysis. They are given by \eqref{K_N=1_Final} and \eqref{K_N=2} for the contributions to the K\"ahler potential from ${\cal N}=1$ sectors and ${\cal N}=2$ sectors, respectively, and by \eqref{tau01N1}-\eqref{tau11N1} for the contributions to the field redefinitions from ${\cal N}=1$ sectors. The ${\cal N}=2$ sector contributions to the field redefinitions, expressed in terms of $\delta E$ and $\overline{G}^{(1)}_{t_i t_j}$, are more complicated and even though our formulas allow us to obtain them for an arbitrary toroidal orientifold, we only worked out the explicit expressions for the $\mathbb{Z}_6'$-orientifold in appendix \ref{redef_Z6'}. 

Thus, our expressions indicate how one can fix the undetermined constants in the formulas for the general structure of the field redefinitions and the K\"ahler potential via string amplitudes. This task is simplified by the fact that the consistency conditions  from supersymmetry and shift symmetry impose relations between different metric components and, thus, between different string amplitudes, cf.\ \eqref{relat-tilde-Gtt} for ${\cal N}=1$ sectors and \eqref{tilde_G_tt_ll}-\eqref{dE/dt_i} for ${\cal N}=2$ sectors. In particular, and interestingly, certain off-diagonal terms of the volume moduli metric have to be non-vanishing at 1-loop level in order to have a non-vanishing contribution to the K\"ahler potential from ${\cal N}=2$ sectors, cf.\ \eqref{K_N=2_KK_0} and \eqref{K_N=2_W_0}. We expect at least some of these contributions to be non-vanishing, namely if the two different volume moduli couple to the same string states, cf.\ figure \ref{KK_winding}. For instance, for the $\mathbb{Z}_6'$-orientifold the $k=3$ sector is formally identical to the $\mathbb{T}^2 \times {\rm K}3$-orientifold discussed in \cite{Antoniadis:1996vw} (and below in appendix \ref{sec:t2k3}), for which it is known that the 1-loop contribution to the K\"ahler potential is non-vanishing. Furthermore, the consistency conditions \eqref{cons-cond1}-\eqref{cons-cond3} (together with our metric ansatz \eqref{Gtt_N=2}) allowed us to fix the 1-loop corrections to the Einstein-frame metric of the axions (the $c$-fields) from those to the Einstein-frame metric of the volume moduli and dilaton. This is remarkable as the computation of the $c$-$c$ amplitudes is very difficult in general. 

We stress again that our analysis shows that it is much simpler to calculate the form of the K\"ahler potential than the form of the field redefinitions. Even though the field redefinitions are indispensable for the derivation of the K\"ahler potential, it is possible to obtain the form of the K\"ahler potential without having to compute the field redefinitions explicitly. For instance in the case of a $\mathbb{Z}_N$-orientifold the complete 1-loop correction to the K\"ahler potential of the diagonal untwisted K\"ahler moduli and the dilaton is given by $K^{(1)} = 8 t_1^2 G_{t_1 t_1}^{(1)}$, which involves the 1-loop correction to the $t_1$-$t_1$ component of the scalar field metric in Einstein-frame, cf.\ \eqref{Ksimple}. Thus, in order to obtain the full 1-loop correction to the K\"ahler potential one only has to calculate $\delta E$ and $\overline{G}^{(1)}_{t_1 t_1}$, cf.\  \eqref{Einstein_frame_Gtt}. This is a huge simplification compared to the eleven different quantities appearing in the metric component $G_{\tau_3 \tau_3}$ for instance, cf.\ \eqref{Gtt_33}, given that it is in general not easy to compute even a single of these quantities. 

There are further interesting directions to pursue in order to generalize or follow up on our results. First of all, even though we focused on 1-loop corrections to the field redefinitions and to the K\"ahler potential, it is an interesting question whether there are already corrections at the level of Euler number $\chi=1$ (i.e.\ from the disk and the projective plane) which do not vanish for vanishing open string scalars. In \cite{Green:2016tfs} an indirect argument based on heterotic-type I duality was given, suggesting a correction to the 4-dimensional Einstein-Hilbert term at this level (i.e.\ a term $\sim e^{\Phi_4} \delta E^{(\chi=1)} R$ in the language of \eqref{1-loop-EF-action-string-1}). After a Weyl transformation this would entail a disk-level contribution to the moduli metric in the Einstein-frame, following the same steps that led from \eqref{1-loop-EF-action-string-1} to \eqref{1-loop-EF-action-2} (with \eqref{Einstein_frame_Gtt} and \eqref{Einstein_frame_Gcc}). In the light of this it would be interesting to revisit the question whether there could also be a correction to the moduli metric in \eqref{1-loop-EF-action-string-1} at $\chi=1$. As mentioned before, naively the momentum expansion of the disk 2-point function of 2 closed string volume moduli seems to indicate that there is none, given that there is no term to quadratic order in the momenta in such an expansion, cf.\ appendix A.2.\ of \cite{Lust:2004cx}. On the other hand, this naive argument would also indicate the absence of a correction to the 4-dimensional Einstein-Hilbert term at disk level (using the momentum expansion of the graviton 2-point function of \cite{Hashimoto:1996bf}), whereas the indirect argument of \cite{Green:2016tfs} seems to suggest the existence of exactly such a term. 

In addition, several generalizations of our method suggest themselves. For instance, it would be interesting to try to incorporate also tree-level $\alpha'$-corrections or corrections from backreaction in case the tadpoles are not cancelled locally. Moreover, in view of potential applications of the 1-loop field redefinitions to the Large Volume scenario, along the lines discussed in \cite{Conlon:2010ji}, it would be worthwhile to incorporate also K\"ahler moduli from the twisted sector (i.e.\ blow-up modes) into the analysis. Finally, it would be interesting to have an independent check for the field redefinitions that we found. For $T_3$ of the $\mathbb{Z}_6'$-orientifold, for instance, this would be feasible by calculating the gauge coupling of the D5-branes, wrapped around the third torus, at Euler number $\chi=-1$ (which is also sometimes called genus $3/2$ order). The gauge kinetic function should be holomorphic in the {\it corrected} variable $T_3$, including the 1-loop correction \eqref{deltatau3}. It should be possible to check this building on earlier work on genus $3/2$ amplitudes, such as \cite{Blau:1987pn,Bianchi:1988fr,Rodrigues:1988bm,Bianchi:1989du,Antoniadis:2004qn,Antoniadis:2005sd}.



\vspace{0.5cm}
\noindent
{\Large{\bf Acknowledgments}}

\noindent
We are very grateful to Marcus Berg, Massimo Bianchi, Kumar Narain, Fernando Quevedo, Ashoke Sen and Stefan Theisen for many extended and enlightening discussions. We also thank Alice Aldi, Ignatios Antoniadis, Ralph Blumenhagen, Ilka Brunner, Subhroneel Chakrabarti, Michele Cicoli, David Ciupke, Atish Dabholkar, Harold Erbin, Edi Gava, Dieter L\"ust, Ruben Minasian, Francisco Morales, Gianfranco Pradisi, Arnab Rudra, Henry Tye, Pierre Vanhove, Erik Verlinde, Giovanni Villadoro and Gianluca Zoccarato for enlightening discussions and Marcus Berg for comments on the draft. The work of M.H.\ was supported by the DFG Transregional Collaborative Research Centre TRR 33 and the Excellence Cluster ``The Origin and the Structure of the Universe'' in Munich and by the German Research Foundation (DFG) within the Emmy-Noether-Program (grant number: HA 3448/3-1). M.H.\ thanks the ICTP in Trieste for hospitality during the beginning of the project.


\begin{appendix}

\section{Change of variables from $t$ to $\tau^{(0)}$}
\label{app:ttau}

In order to perform the change of variables from $t$ to $\tau^{(0)}$, it is convenient to introduce two sets of coordinates as 
\be
& x_0= 2 t_0,\quad  x_i = \ln t_i, \qquad i \in \{1, 2, 3 \} \,,\\
& y_i=  \ln \tau_i^{(0)} , \qquad i \in \{0, 1, 2, 3 \}  \,.
\ee
Using \eqref{t0-tau} and \eqref{t-tau0} one easily verifies that the relation between the $x$ and $y$ coordinates is linear, 
\be \label{xAy}
\begin{pmatrix} x_0 \\ x_1 \\ x_2 \\ x_3
\end{pmatrix}= A 
\begin{pmatrix} y_0 \\ y_1 \\ y_2 \\ y_3
\end{pmatrix}
\ee
with $A$ being a constant orthogonal matrix given by \eqref{A}.
Note that 
\be
A^{T} A = I_4 \,,
\ee
where $I_4$ is the $4$-dimensional identity matrix.

The kinetic terms of the $t$-moduli in the Einstein-frame action \eqref{1-loop-EF-action-2} can be expressed in terms of $x$ and $y$ coordinates (we use the compact notation $dx_i dx_j = -\partial_\mu x_i \partial^{\mu} x_j$) as 
\be
&& 
- \sum_{i,j=0}^3 \left[ G^{(0)}_{t_i t_j} +  G^{(1)}_{t_i t_j} \right] \partial_\mu t_i \partial^\mu t_j \nonumber \\
&=& 
  \left[\frac{1 +  G^{(1)}_{t_0 t_0}}{4} \right] d (2 t_0) d (2 t_0) 
+ 2 \sum_{i=1}^3  \left(\frac{t_i G^{(1)}_{t_0 t_i}}{2}\right) d (2 t_0) d (\ln t_i) 
+ \sum_{i,j=1}^3 t_i t_j \left[ G^{(0)}_{t_i t_j} +  G^{(1)}_{ij} \right] d(\ln t_i) d(\ln t_j) \nonumber \\
&=& 
  \left[\frac{1 +  G^{(1)}_{t_0 t_0}}{4} \right] d x_0 d x_0 
+ 2 \sum_{i=1}^3  \left(\frac{t_i G^{(1)}_{t_0 t_i}}{2}\right) d x_0 d x_i 
+ \sum_{i,j=1}^3 t_i t_j \left[ G^{(0)}_{t_i t_j} +  G^{(1)}_{t_i t_j} \right] dx_i d x_j  \nonumber \\
&\equiv&
\sum_{i,j=0}^3 X_{i j} dx_i dx_j  =\sum_{i,j=0}^3 \left(X^{(0)}_{i j}+ X^{(1)}_{i j}\right) dx_i dx_j \nonumber  \\
&=& \sum_{i,j=0}^3 \left(A^{T} X A\right)_{i j} dy_i dy_j =\sum_{i,j=0}^3 \left[A^{T} \left(X^{(0)}+ X^{(1)}\right) A\right]_{i j} dy_i dy_j  \nonumber \\
&\equiv&
\sum_{i,j=0}^3 Y_{i j} dy_i dy_j = \sum_{i,j=0}^3 \left(Y^{(0)}_{i j}+ Y^{(1)}_{i j}\right) dy_i dy_j  \nonumber \\
&=& \sum_{i,j=0}^3 \left(\frac{Y_{i j}}{\tau_i^{(0)} \tau_j^{(0)}}\right) d\tau_i^{(0)} d\tau_j^{(0)} = \sum_{i,j=0}^3 
\left(\frac{Y^{(0)}_{i j}}{\tau_i^{(0)} \tau_j^{(0)}}+ \frac{Y^{(1)}_{i j}}{\tau_i^{(0)} \tau_j^{(0)}}\right) d\tau_i^{(0)} d\tau_j^{(0)}  \nonumber \\
&\equiv& \sum_{i,j=0}^3 G_{\tau_i^{(0)} \tau_j^{(0)}} d\tau_i^{(0)} d\tau_j^{(0)} =
 \sum_{i,j=0}^3 \left(G_{\tau_i^{(0)} \tau_j^{(0)}}^{(0)}+ G_{\tau_i^{(0)} \tau_j^{(0)}}^{(1)} \right) d\tau_i^{(0)} d\tau_j^{(0)} \,. \label{changettau}
\ee
Here
\be
X_{0 0} = \frac{1 +  G^{(1)}_{t_0 t_0}}{4} \,, \qquad
X_{0 i}= \frac{t_i G^{(1)}_{t_0 t_i}}{2}\,, \qquad
X_{i j} = t_i t_j \left[ G^{(0)}_{t_i t_j} +  G^{(1)}_{t_i t_j} \right] \qquad \text{with $i,j \in \{1,2,3\}$}
\ee
and (using $G^{(0)}_{t_i t_j} = \delta_{ij}/(4 t_i^2)$ for $i,j \in \{1,2,3\}$)
\be
X^{(0)} = \frac{I_4}{4} \qquad , \qquad 
X^{(1)} =
\begin{pmatrix}
\frac{  G^{(1)}_{t_0 t_0}}{4}, & \frac{t_1 G^{(1)}_{t_0 t_1}}{2}, & \frac{t_2 G^{(1)}_{t_0 t_2}}{2}, & \frac{t_3 G^{(1)}_{t_0 t_3}}{2} \\
\frac{t_1 G^{(1)}_{t_0 t_1}}{2}, & t_1^2 G^{(1)}_{t_1 t_1}, &  t_1 t_2 G^{(1)}_{t_1 t_2}, &  t_1 t_3 G^{(1)}_{t_1 t_3} \\
\frac{t_2 G^{(1)}_{t_0 t_2}}{2}, &  t_1 t_2 G^{(1)}_{t_1 t_2},  & t_2^2 G^{(1)}_{t_2 t_2}, & t_2 t_3 G^{(1)}_{t_2 t_3} \\
\frac{t_3 G^{(1)}_{t_0 t_3}}{2}, & t_1 t_3 G^{(1)}_{t_1 t_3},  & t_2 t_3 G^{(1)}_{t_2 t_3}, & t_3^2 G^{(1)}_{t_3 t_3}
\end{pmatrix} \,, 
\ee
\be
Y^{(0)} = A^T X^{(0)} A = A^T \left(\frac{I_4}{4}\right) A= \frac{I_4}{4} =X^{(0)} \qquad , \qquad Y^{(1)} = A^T X^{(1)} A \,.
\ee

Concretely $Y_{ij}^{(1)}=\left(A^T X^{(1)} A\right)_{ij}$ reads
\be
Y_{00}^{(1)} &=& \frac{1}{4} \left(\frac{G_{t_0 t_0}^{(1)}}{4} - t_1 G_{t_0 t_1}^{(1)} -t_2 G_{t_0 t_2}^{(1)} -t_3 G_{t_0 t_3}^{(1)} + 
t_1^2  G_{t_1 t_1}^{(1)} +t_2^2  G_{t_2 t_2}^{(1)}+t_3^2  G_{t_3 t_3}^{(1)} +  \right. \nonumber \\
&& \left. \qquad \quad  + 2 t_1 t_2  G_{t_1 t_2}^{(1)} + 2 t_1 t_3  G_{t_1 t_3}^{(1)} + 2 t_2 t_3  G_{t_2 t_3}^{(1)}   \right) \ ,   \label{Y00} \\
Y_{01}^{(1)} &=& \frac{1}{4} \left(\frac{G_{t_0 t_0}^{(1)}}{4} - t_1 G_{t_0 t_1}^{(1)}  + 
t_1^2  G_{t_1 t_1}^{(1)} - t_2^2  G_{t_2 t_2}^{(1)}- t_3^2  G_{t_3 t_3}^{(1)} - 2 t_2 t_3 G_{t_2 t_3}^{(1)}  \right)\ , \label{Y01} \\
Y_{02}^{(1)} &=& \frac{1}{4} \left(\frac{G_{t_0 t_0}^{(1)}}{4} -t_2 G_{t_0 t_2}^{(1)} - 
t_1^2  G_{t_1 t_1}^{(1)} +t_2^2  G_{t_2 t_2}^{(1)}-t_3^2  G_{t_3 t_3}^{(1)}  - 2 t_1 t_3 G_{t_1 t_3}^{(1)} \right) \ ,  \label{Y02} \\
Y_{03}^{(1)} &=& \frac{1}{4} \left(\frac{G_{t_0 t_0}^{(1)}}{4} -t_3 G_{t_0 t_3}^{(1)} - 
t_1^2  G_{t_1 t_1}^{(1)} - t_2^2  G_{t_2 t_2}^{(1)}+t_3^2  G_{t_3 t_3}^{(1)}  - 2 t_1 t_2 G_{t_1 t_2}^{(1)} \right) \ ,  \label{Y03} \\
Y_{11}^{(1)} &=& \frac{1}{4} \left(\frac{G_{t_0 t_0}^{(1)}}{4} - t_1 G_{t_0 t_1}^{(1)} +t_2 G_{t_0 t_2}^{(1)} +t_3 G_{t_0 t_3}^{(1)} + 
t_1^2  G_{t_1 t_1}^{(1)} +t_2^2  G_{t_2 t_2}^{(1)}+t_3^2  G_{t_3 t_3}^{(1)}  \right. \nonumber \\
&& \left. \qquad \quad  - 2 t_1 t_2  G_{t_1 t_2}^{(1)} - 2 t_1 t_3  G_{t_1 t_3}^{(1)} + 2 t_2 t_3  G_{t_2 t_3}^{(1)}   \right) \ ,   \label{Y11} \\
Y_{12}^{(1)} &=& \frac{1}{4} \left(\frac{G_{t_0 t_0}^{(1)}}{4}  + t_3 G_{t_0 t_3}^{(1)} - 
t_1^2  G_{t_1 t_1}^{(1)} - t_2^2  G_{t_2 t_2}^{(1)}+t_3^2  G_{t_3 t_3}^{(1)}  + 2 t_1 t_2 G_{t_1 t_2}^{(1)}  \right) \ ,  \label{Y12} \\
Y_{13}^{(1)} &=& \frac{1}{4} \left(\frac{G_{t_0 t_0}^{(1)}}{4} + t_2 G_{t_0 t_2}^{(1)} - 
t_1^2  G_{t_1 t_1}^{(1)} + t_2^2  G_{t_2 t_2}^{(1)}- t_3^2  G_{t_3 t_3}^{(1)}  + 2 t_1 t_3 G_{t_1 t_3}^{(1)} \right) \ ,  \label{Y13} \\
Y_{22}^{(1)} &=& \frac{1}{4} \left(\frac{G_{t_0 t_0}^{(1)}}{4} + t_1 G_{t_0 t_1}^{(1)} -t_2 G_{t_0 t_2}^{(1)} + t_3 G_{t_0 t_3}^{(1)} + 
t_1^2  G_{t_1 t_1}^{(1)} +t_2^2  G_{t_2 t_2}^{(1)}+t_3^2  G_{t_3 t_3}^{(1)}  \right. \nonumber \\
&& \left. \qquad \quad  - 2 t_1 t_2  G_{t_1 t_2}^{(1)} + 2 t_1 t_3  G_{t_1 t_3}^{(1)} - 2 t_2 t_3  G_{t_2 t_3}^{(1)}   \right)\ ,  \label{Y22} \\
Y_{23}^{(1)} &=& \frac{1}{4} \left(\frac{G_{t_0 t_0}^{(1)}}{4} + t_1 G_{t_0 t_1}^{(1)} + 
t_1^2  G_{t_1 t_1}^{(1)} -t_2^2  G_{t_2 t_2}^{(1)}-t_3^2  G_{t_3 t_3}^{(1)} + 2 t_2 t_3 G_{t_2 t_3}^{(1)}  \right) \ ,  \label{Y23} \\
Y_{33}^{(1)} &=& \frac{1}{4} \left(\frac{G_{t_0 t_0}^{(1)}}{4} + t_1 G_{t_0 t_1}^{(1)} + t_2 G_{t_0 t_2}^{(1)} -t_3 G_{t_0 t_3}^{(1)} + 
t_1^2  G_{t_1 t_1}^{(1)} +t_2^2  G_{t_2 t_2}^{(1)}+t_3^2  G_{t_3 t_3}^{(1)}  \right. \nonumber \\ 
&& \left. \qquad \quad  + 2 t_1 t_2  G_{t_1 t_2}^{(1)} - 2 t_1 t_3  G_{t_1 t_3}^{(1)} - 2 t_2 t_3  G_{t_2 t_3}^{(1)} \right)  \ .\label{Y33}
\ee
From here we easily verify
\be
Y_{00}^{(1)}+Y_{11}^{(1)}-2Y_{01}^{(1)}&=& t_2^2 G_{t_2 t_2}^{(1)}+t_3^2 G_{t_3 t_3}^{(1)} + 2 t_2 t_3 G_{t_2 t_3}^{(1)} \ ,\label{Y0-1} \\ 
Y_{22}^{(1)}+Y_{33}^{(1)}-2Y_{23}^{(1)} &=& t_2^2 G_{t_2 t_2}^{(1)}+t_3^2 G_{t_3 t_3}^{(1)} - 2 t_2 t_3 G_{t_2 t_3}^{(1)} \ ,\label{Y2-3} \\
Y_{00}^{(1)}+Y_{22}^{(1)}-2Y_{02}^{(1)} &=& t_1^2 G_{t_1 t_1}^{(1)}+t_3^2 G_{t_3 t_3}^{(1)} + 2 t_1 t_3 G_{t_1 t_3}^{(1)} \ , \label{Y0-2} \\
Y_{11}^{(1)}+Y_{33}^{(1)}-2Y_{13}^{(1)} &=& t_1^2 G_{t_1 t_1}^{(1)}+t_3^2 G_{t_3 t_3}^{(1)} - 2 t_1 t_3 G_{t_1 t_3}^{(1)} \ , \label{Y1-3} \\
Y_{00}^{(1)}+Y_{33}^{(1)}-2Y_{03}^{(1)} &=& t_1^2 G_{t_1 t_1}^{(1)}+t_2^2 G_{t_2 t_2}^{(1)}  + 2 t_1 t_2 G_{t_1 t_2}^{(1)} \,, \label{Y0-3} \\
Y_{11}^{(1)}+Y_{22}^{(1)}-2Y_{12}^{(1)} &=& t_1^2 G_{t_1 t_1}^{(1)}+t_2^2 G_{t_2 t_2}^{(1)}  - 2 t_1 t_2 G_{t_1 t_2}^{(1)}  \,.\label{Y1-2}
\ee    


\section{Moduli dependence of $G_{c_i^{(0)} c_j^{(0)}}^{(1)}$ in $\mathcal{N}=2$ sectors \label{app_beta_ij}}

Here we prove \eqref{Gcc_N=2} from \eqref{Gtt_N=2}, 
using the constraints \eqref{cons-cond1} and \eqref{cons-cond3}. 
To this end, we start with a more general form of \eqref{Gtt_N=2} and \eqref{Gcc_N=2} (with the replacements $\tau_i^{(0)} \to \tau_i$, cf. footnote \ref{interchange}) as\footnote{ By dilaton counting for 1-loop corrections in comparison with the tree-level metric ($\sim \mathcal{O}\left(1/\tau^2\right)$) one sees that $\hat{\alpha}_{i j}$ and $\hat{\beta}_{i j}$ in \eqref{Gtt-app}-\eqref{Gcc-app} 
can not depend on the dilaton $t_0$, but can depend at most on $t_i$ (with $i \in \{1,2,3\}$) and the complex-structure moduli $U$.  \label{app-beta}}
\be
G_{\tau_i^{(0)} \tau_j^{(0)}}^{(1)}&=&\frac{\hat{\alpha}_{i j}(U, t)}{\tau_i \tau_j \sqrt{\tau_0 \tau_1 \tau_2 \tau_3}} \ , \label{Gtt-app} \\
G_{c_i^{(0)} c_j^{(0)}}^{(1)}&=&\frac{\hat{\beta}_{i j}(U, t)}{\tau_i \tau_j \sqrt{\tau_0 \tau_1 \tau_2 \tau_3}} \ . \label{Gcc-app}
\ee
Let us plug these expressions into the constraints \eqref{cons-cond1}-\eqref{cons-cond3}, which then reduce to ($i \neq j \neq k$)
\be
2 \hat{\alpha}_{i j} = (D_i+D_j)\, \hat{\beta}_{i j} - D_j \,\hat{\beta}_{ii} - D_i\, \hat{\beta}_{j j}  \,, \label{cond1-app}
\ee
\be
D_k\, \hat{\beta}_{i j}  \,=\, D_i\, \hat{\beta}_{j k}  \,=\, D_j \,\hat{\beta}_{k i} \,, \label{cond2-app}
\ee
\be
D_i \,D_j \,\hat{\beta}_{i i}  = \left(D_i^2 -1 \right) \hat{\beta}_{i j}  - D_j \,\hat{\alpha}_{ii} \,,\label{cond3-app}
\ee
respectively.  Here we introduced the operator $D_i$ which is defined as
\be
D_i \equiv -\frac{1}{2}+\tau_i \, \frac{\partial}{\partial \tau_i}\,. \label{D_i} 
\ee
Note that the operator $D_i$ given in \eqref{D_i} preserves the form of any power function of the $\tau$-variables that is acted on by $D_i$. This is to say that a power function of $\tau$s is an eigenfunction of the $D_i$ operators and different power functions do not mix with each other under the action of $D_i$ operators. (Note that the eigenvalue might be zero, which is the case for instance for $\tau_i^{1/2}$ since $D_i \tau_i^{1/2}=0$.) Equations \eqref{cond1-app} and \eqref{cond3-app} can be used to derive
\be
(D_i +D_j ) \, \hat{\beta}_{i j} &=& 2 D_i \, D_j \, \hat{\alpha}_{i j} -D_j^2\, \hat{\alpha}_{i i} -D_i^2\, \hat{\alpha}_{j j}  \qquad \qquad \text{for} \quad  i\neq j \ . \label{beta_ij-app} 
\ee
This follows by acting with $D_i D_j$ on \eqref{cond1-app} and then using \eqref{cond3-app} acted on by $D_j$ 
(and also the expression obtained by interchanging $i \leftrightarrow j$ in the latter). 

Now we specify the form of $\hat{\alpha}_{ij}$ in \eqref{Gtt-app} according to our ansatz \eqref{Gtt_N=2} in the main text, i.e.\ (for arbitrary $i$ and $j$)
\be
\hat{\alpha}_{ij}(U, t)& = & \sum_{m=\pm 1} \sum_{l=1}^3 
\alpha_{i j}^{(m,l)}(U_l) \cdot t_l^m  \nonumber \\
&=&  \sum_{m=\pm 1} \sum_{l=1}^3 
\alpha_{i j}^{(m,l)}(U_l) \cdot  \left(\frac{\tau_0\tau_l}{\tau_I \tau_J }\right)^{m/2} 
\qquad \text{with} \quad I\neq J \in \{1,2,3\} \setminus \{l\} \,. \label{dec-alpha-app}
\ee 
In the second line we expressed $t_l$  in terms of $\tau$ using \eqref{t-tau0} (again replacing $\tau^{(0)} \to \tau$ according to footnote \ref{interchange}). We want to use \eqref{dec-alpha-app} in \eqref{beta_ij-app} and \eqref{cond3-app} in order to show that their solutions $\hat{\beta}$ are of the same form as \eqref{dec-alpha-app},
i.e.\ (for arbitrary $i$ and $j$)
\be
\hat{\beta}_{ij}=  \sum_{m=\pm 1} \sum_{l=1}^3 
\beta_{i j}^{(m,l)}(U_l) \cdot t_l^m  \ . \label{dec-beta-app}
\ee
This then shows that our ansatz \eqref{Gcc_N=2} follows directly from \eqref{Gtt_N=2} once the constraints  \eqref{cons-cond1} and \eqref{cons-cond3} are imposed. 

To this end we first use the fact that the string 1-loop corrections from ${\cal N}=2$ sectors can only depend non-trivially on the volume and complex structure of the torus along which the KK/winding sum arises (up to the trivial moduli dependence from the loop counting factor and the normalization factors of the vertex operators). Thus, we have (for arbitrary $i$ and $j$)
\be
\hat{\beta}_{ij}(U, t) =  \sum_{l=1}^3 \hat{\beta}_{i j}^{(l)}(U_l, t_l) \ . \label{beta_ij_l}
\ee  
Note that 
\be
D_i \, f(t_l)= -\frac{f(t_l)}{2}  + 
A_{li} \frac{\partial\, f(t_l)}{\partial \left(\ln t_l\right)} \,, \label{D-f(t)}
\ee
where we used the chain rule 
\be
\frac{\partial f(t_l)}{\partial (\ln\tau_i)} = 
A_{li} \frac{\partial f(t_l)}{\partial \left(\ln t_l\right)} \,, \label{D-beta}
\ee 
with
\be
A_{li} = \frac{\partial (\ln t_l)}{\partial(\ln \tau_i)} 
\ee
being the $(l,i)$ element of matrix $A$ given in \eqref{A}, cf.\ \eqref{xAy}. (Note that $l \in \{1,2,3\}$ and $i \in \{0, \ldots ,3\}$.)

Now we solve \eqref{beta_ij-app} to obtain the off-diagonal $\hat{\beta}_{ij}$. First, from the property of the $D_i$ operators mentioned below \eqref{D_i} one can easily verify that \eqref{beta_ij-app} has a particular solution of the form \eqref{dec-beta-app} (this also requires some relations between the $\beta_{i j}^{(m,l)}$ and $\alpha_{i j}^{(m,l)}$, which are discussed in the main text, see section \ref{consistency_N2}).  On the other hand, \eqref{beta_ij-app} is a linear inhomogeneous differential equation and its general solution is given by the sum of a particular solution and the general solution to the homogeneous equation. Using \eqref{beta_ij_l} and \eqref{D-f(t)} the homogeneous part of \eqref{beta_ij-app} is given by 
\be
(A_{li} + A_{lj})\, t_l\, \frac{\partial \hat{\beta}_{i j}^{(l)}(t_l)}{\partial t_l} = \hat{\beta}_{i j}^{(l)}(t_l) \qquad \text{for} \quad i\neq j\ .
\ee
This equation holds for all $l$ separately. It is easy to see, using \eqref{A}, that this has the solutions $\hat{\beta}_{i j}^{(l)}(t_l) = c\, t_l^{\pm1}$ for $A_{li} + A_{lj}=\pm1$ (and $\hat{\beta}_{i j}^{(l)}(t_l) = 0$ for $A_{li} + A_{lj}=0$) with $c$ being an integration constant. This shows that the ansatz \eqref{dec-beta-app}, for $i \neq j$, is sufficient to describe the general solution of the constraint \eqref{beta_ij-app}.

Now let us turn to the diagonal components $\hat{\beta}_{ii}$ which are constrained by \eqref{cond3-app}. 
We follow the same steps as above for the off-diagonal components, using \eqref{dec-beta-app} for $\hat{\beta}_{ij}$ (with $i \neq j$) and \eqref{beta_ij_l} for $\hat{\beta}_{ii}$. It is easy to see that the particular solution of the inhomogeneous equation \eqref{cond3-app} is of the form \eqref{dec-beta-app}. Now we solve the homogeneous part of \eqref{cond3-app}, i.e.\ $D_i \,D_j \,\hat{\beta}_{i i}  =0$. Using \eqref{beta_ij_l} and \eqref{D-f(t)} this can be written as
\be
4 \, t_l^2 \frac{\partial^2 \hat{\beta}_{ii}^{(l)}}{\partial t_l^2} A_{l i} A_{l j} + t_l \frac{\partial \hat{\beta}_{ii}^{(l)}}{\partial t_l} (4 A_{l i} A_{l j}-2A_{l i}- 2A_{l j}) 
+ \hat{\beta}_{ii}^{(l)} =0 \qquad \text{for} \quad i \neq j \ , \label{hom-ii}
\ee
which again holds for each $l$ separately. Using \eqref{A}, one can see that the possible solutions of the above homogeneous equation are $\hat{\beta}_{ii}^{(l)}(t_l)=\big\{ t_l(c_1 +c_2 \log t_l), \, t_l^{-1}(c_3 +c_4 \log t_l), \, c_5\, t_l^{-1}+ c_6\, t_l \big\}$ for $(A_{li}, A_{lj})= \big\{(\frac{1}{2},\frac{1}{2}), (-\frac{1}{2},-\frac{1}{2}), {\rm otherwise} \big\}$, respectively, where $c_1, \ldots , c_6$ are integration constants. Now for fixed indices $l$ and $i$, the value of $A_{li}$ is fixed. Then we are free to choose the index $j$ in such a way that either $A_{lj}=1/2$ or $-1/2$ in \eqref{hom-ii}, and both values of $A_{lj}$ should give the same solution $\hat{\beta}_{ii}^{(l)}(t_l)$ since \eqref{hom-ii} should hold for any $j$ different from $i$. If $A_{li}=1/2$, there are choices for the index $j$ ($\neq i$) such that $(A_{li}, A_{lj})= (\frac{1}{2},\frac{1}{2})$ or $(A_{li}, A_{lj})=(\frac{1}{2},-\frac{1}{2})$. Thus, we obtain $\hat{\beta}_{ii}^{(l)}(t_l)= t_l(c_1 +c_2 \log t_l) =c_5\, t_l^{-1}+ c_6\, t_l $. This implies $c_1=c_6$ and $c_2=0=c_5$, so that we obtain $\hat{\beta}_{ii}^{(l)}(t_l)=c\, t_l$ when $A_{li}=1/2$. Analogously one can obtain  $\hat{\beta}_{ii}^{(l)}(t_l)=c\, t_l^{-1}$ when $A_{li}=-1/2$. Again we see that the ansatz \eqref{dec-beta-app}, for $i = j$, is sufficient to describe the general solution of the constraint \eqref{cond3-app}. Thus we have proved \eqref{dec-beta-app} (i.e.\ \eqref{Gcc_N=2}) for arbitrary $i$ and $j$.


\section{Details of calculations for ${\cal N}=2$ sectors}
\label{detailsN2}

In this appendix we fill in some calculational details that we left out in the derivation of \eqref{tau_a^(-1,l)}-\eqref{tau_I^(1,l)} and \eqref{K_N=2_KK}-\eqref{K_N=2_W}.

Let us begin with the derivation of the field redefinitions. Consider \eqref{eq-ii}, which reads for the $(-1,l)$- and $(1,l)$-sectors
\be
\frac{\partial \left(\tau_i^{(1)|(-1,l)}(\tau)\right)}{\partial \tau_i}
&=&\frac{2 \left(\alpha_{ii}^{(-1,l)} -\beta_{ii}^{(-1,l)}\right)}{
\tau_0 \tau_l}\ , \\
\frac{\partial \left(\tau_i^{(1)|(1,l)}(\tau)\right)}{\partial \tau_i}
&=&\frac{2 \left(\alpha_{ii}^{(1,l)} -\beta_{ii}^{(1,l)}\right) \tau_l}{
\tau_1 \tau_2 \tau_3}\ ,
\ee 
where we used \eqref{N=2_Gtautau_0} and \eqref{N=2_Gcc_0}.
Splitting the indices according to \eqref{a-b} and \eqref{I-J} gives
\be
\frac{\partial \left(\tau_a^{(1)|(-1,l)}(\tau)\right)}{\partial \tau_a}&=& 
\frac{2 \left(\alpha_{aa}^{(-1,l)} -\beta_{aa}^{(-1,l)}\right) }{
 \tau_0 \tau_l }=0 \,,\\
\frac{\partial \left(\tau_I^{(1)|(-1,l)}(\tau)\right)}{\partial \tau_I}&=&
\frac{2 \left(\alpha_{II}^{(-1,l)} -\beta_{II}^{(-1,l)}\right) }{
 \tau_0 \tau_l }=  \frac{-4 \alpha_{aI}^{(-1,l)}}{
 \tau_0 \tau_l } \,, \\
\frac{\partial \left(\tau_a^{(1)|(1,l)}(\tau)\right)}{\partial \tau_a}&=& 
\frac{2 \tau_l \left(\alpha_{aa}^{(1,l)} -\beta_{aa}^{(1,l)}\right) }{
 \tau_1 \tau_2 \tau_3 }= \frac{-4 \alpha_{aI}^{(1,l)}}{
 \tau_I \tau_J} \,, \label{dtauataua}\\
\frac{\partial \left(\tau_I^{(1)|(1,l)}(\tau)\right)}{\partial \tau_I}&=&
\frac{2 \tau_l \left(\alpha_{II}^{(1,l)} -\beta_{II}^{(1,l)}\right) }{
 \tau_1 \tau_2 \tau_3 }=0 \,,
\ee
where $I \neq J$ in \eqref{dtauataua} and we used \eqref{KK-beta-alpha_aa}, \eqref{KK-beta-alpha_II} and \eqref{KK-alpha_II}, \eqref{W-beta-alpha_aa} and \eqref{W-alpha_aa} and \eqref{W-beta-alpha_II}, respectively. These equations have the solutions
\be
\tau_a^{(1)|(-1,l)}(\tau) &=& C_a^{(-1,l)}(\hat{\tau}_a) \,, \label{taua-1} \\
\tau_I^{(1)|(-1,l)}(\tau) &=&  C_I^{(-1,l)}(\hat{\tau}_I)- 
\frac{4 \tau_I \alpha_{aI}^{(-1,l)}}{
 \tau_0 \tau_l } \,, \label{tauI-1}\\
\tau_a^{(1)|(1,l)}(\tau) &=& C_a^{(1,l)}(\hat{\tau}_a) -
\frac{4 \tau_a \alpha_{aI}^{(1,l)}}{
 \tau_I \tau_J} \,,\label{taua1} \\
\tau_I^{(1)|(1,l)}(\tau) &=& C_I^{(1,l)}(\hat{\tau}_I) \label{tauI1} \ .
\ee
Here again $I \neq J$ in \eqref{taua1} and $C_i^{(m,l)}(\hat{\tau}_i)$ are ``integration" constants with respect to $\tau_i$ (i.e. $\frac{\partial C_i^{(m,l)}}{\partial \tau_i}=0$). They are determined from \eqref{eq-ij}. For the terms arising from closed string winding states, \eqref{eq-ij} leads to
\be
\frac{\partial \left( \tau_a^{(1)|{(-1,l)}}\right)}{\partial \tau_b}
&=&2 \tau_a^3\left[ \partial_{\tau_a}\left(G_{c_a^{(0)} c_b^{(0)}}^{(1)|{(-1,l)}}(\tau)\right) - 
\partial_{\tau_b}\left(G_{c_a^{(0)} c_a^{(0)}}^{(1)|{(-1,l)}}(\tau)\right)\right] \label{dtauataub-1} \\
&=& \frac{1}{\tau_b^2}
\left[-4 \beta_{ab}^{(-1,l)} + 2 \beta_{aa}^{(-1,l)} \right] = \frac{1}{\tau_b^2} \left[-2 \alpha_{bb}^{(-1,l)} + 4 \alpha_{ab}^{(-1,l)} \right] \,,  \\
\frac{\partial \left(\tau_a^{(1)|{(-1,l)}}\right)}{\partial \tau_I}
&=&2 \tau_a^3\left[ \partial_{\tau_a}\left(G_{c_a^{(0)} c_I^{(0)}}^{(1)|{(-1,l)}}(\tau)\right) - 
\partial_{\tau_I}\left(G_{c_a^{(0)} c_a^{(0)}}^{(1)|{(-1,l)}}(\tau)\right)\right]\\
&=&\frac{1}{\tau_I \tau_b} \left[-4 \beta_{aI}^{(-1,l)} \right] = 0 \,,\\
\frac{\partial \left( \tau_I^{(1)|{(-1,l)}}\right)}{\partial \tau_a}
&=&2 \tau_I^3\left[ \partial_{\tau_I}\left(G_{c_I^{(0)} c_a^{(0)}}^{(1)|{(-1,l)}}(\tau)\right) - 
\partial_{\tau_a}\left(G_{c_I^{(0)} c_I^{(0)}}^{(1)|{(-1,l)}}(\tau)\right)\right]\\
&=& \frac{\tau_I}{\tau_a^2 \tau_b} \left[-2 \beta_{aI}^{(-1,l)} + 2 \beta_{II}^{(-1,l)} \right] =  \frac{\tau_I}{\tau_a^2 \tau_b} \left[4 \alpha_{aI}^{(-1,l)} \right] \,,\\
\frac{\partial \left( \tau_I^{(1)|{(-1,l)}}\right)}{\partial \tau_J}
&=&2 \tau_I^3\left[ \partial_{\tau_I}\left(G_{c_I^{(0)} c_J^{(0)}}^{(1)|{(-1,l)}}(\tau)\right) - 
\partial_{\tau_J}\left(G_{c_I^{(0)} c_I^{(0)}}^{(1)|{(-1,l)}}(\tau)\right)\right]  \\
&=& \frac{\tau_I}{\tau_J \tau_0 \tau_l} \left[-2 \beta_{IJ}^{(-1,l)}\right] = 0\,.
\ee
Here $a \neq b$ and  $I \neq J$ and we used \eqref{KK-beta-alpha_aa} - \eqref{KK-alpha_IJ}. Moreover, for the contributions from closed string Kaluza-Klein modes we have
\be
\frac{\partial \left( \tau_a^{(1)|{(1,l)}}\right)}{\partial \tau_b}
&=&2 \tau_a^3\left[ \partial_{\tau_a}\left(G_{c_a^{(0)} c_b^{(0)}}^{(1)|{(1,l)}}(\tau)\right) - 
\partial_{\tau_b}\left(G_{c_a^{(0)} c_a^{(0)}}^{(1)|{(1,l)}}(\tau)\right)\right] \\
&=& \left(\frac{\tau_a}{\tau_b}\right)
\left(\frac{ \tau_l} { \tau_1 \tau_2 \tau_3}\right) \left[-2 \beta_{ab}^{(1,l)} \right] = 0 \,,\\
\frac{\partial \left(\tau_a^{(1)|{(1,l)}}\right)}{\partial \tau_I}
&=&2 \tau_a^3\left[ \partial_{\tau_a}\left(G_{c_a^{(0)} c_I^{(0)}}^{(1)|{(1,l)}}(\tau)\right) - 
\partial_{\tau_I}\left(G_{c_a^{(0)} c_a^{(0)}}^{(1)|{(1,l)}}(\tau)\right)\right]\\
&=&\left(\frac{\tau_a}{\tau_I}\right)
\left(\frac{ \tau_l} { \tau_1 \tau_2 \tau_3}\right) \left[-2 \beta_{aI}^{(1,l)} + 2 \beta_{aa}^{(1,l)}\right] = \frac{\tau_a}{\tau_I^2 \tau_J} \left[4 \alpha_{aI}^{(1,l)}\right] \,,\\
\frac{\partial \left( \tau_I^{(1)|{(1,l)}}\right)}{\partial \tau_a}
&=&2 \tau_I^3\left[ \partial_{\tau_I}\left(G_{c_I^{(0)} c_a^{(0)}}^{(1)|{(1,l)}}(\tau)\right) - 
\partial_{\tau_a}\left(G_{c_I^{(0)} c_I^{(0)}}^{(1)|{(1,l)}}(\tau)\right)\right]\\
&=& \left(\frac{\tau_I}{\tau_a}\right)
\left(\frac{ \tau_l} { \tau_1 \tau_2 \tau_3}\right) \left[-4 \beta_{aI}^{(1,l)}\right] = 0\,,\\
\frac{\partial \left( \tau_I^{(1)|{(1,l)}}\right)}{\partial \tau_J}
&=&2 \tau_I^3\left[ \partial_{\tau_I}\left(G_{c_I^{(0)} c_J^{(0)}}^{(1)|{(1,l)}}(\tau)\right) - 
\partial_{\tau_J}\left(G_{c_I^{(0)} c_I^{(0)}}^{(1)|{(1,l)}}(\tau)\right)\right]  \\
&=& \left(\frac{\tau_I}{\tau_J}\right)
\left(\frac{ \tau_l} { \tau_1 \tau_2 \tau_3}\right) \left[-4 \beta_{IJ}^{(1,l)} + 2 \beta_{II}^{(1,l)} \right] = \frac{1} {\tau_J^2} \left[ - 2 \alpha_{JJ}^{(1,l)} + 4 \alpha_{IJ}^{(1,l)} \right]\,. \label{dtauItauJ1}
\ee
Again $a \neq b$ and  $I \neq J$ and we used \eqref{W-beta-alpha_II} - \eqref{W-alpha_ab}. Comparing \eqref{taua-1}-\eqref{tauI1} with \eqref{dtauataub-1}-\eqref{dtauItauJ1} we obtain
\be
C_a^{(-1,l)}&=& \frac{2\left(\alpha_{bb}^{(-1,l)} - 2 \alpha_{ab}^{(-1,l)}\right)}{\tau_b}\ , \label{Caminus1} \\
C_I^{(-1,l)}&=&0\ , \label{CIminus1} \\
C_a^{(1,l)}&=&0\ , \label{Ca1}\\
C_I^{(1,l)}&=& \frac{2\left(\alpha_{JJ}^{(1,l)} - 2 \alpha_{IJ}^{(1,l)}\right)}{\tau_J} \label{CI1}
\ee 
with $a \neq b$ and $I \neq J$.\footnote{Of course, we could add arbitrary constants to \eqref{Caminus1}-\eqref{CI1}, but these would just amount to holomorphic field redefinitions.}  Plugging this into \eqref{taua-1}-\eqref{tauI1} leads to the formulas for the field redefinitions given in the main text, i.e.\ \eqref{tau_a^(-1,l)} - \eqref{tau_I^(1,l)}.

For the K\"ahler potential, let us start with the diagonal components of the K\"ahler metric. Using \eqref{K2_N=2} together with \eqref{tau_a^(-1,l)}-\eqref{tau_I^(1,l)}, \eqref{KK-beta-alpha_aa}-\eqref{W-alpha_ab} and \eqref{N=2_Gcc_0}, we obtain the conditions
\be
\frac{1}{4}\,   \frac{\partial^2 K^{(1)|(-1,l)}(\tau)}{\partial \tau_a \partial \tau_a} &=&  
 \frac{\alpha_{aa}^{(-1,l)}+\alpha_{bb}^{(-1,l)} - 2 \alpha_{ab}^{(-1,l)}}{\tau_a^3 \tau_b}\ ,  \\
\frac{1}{4}\,   \frac{\partial^2 K^{(1)|(-1,l)}(\tau)}{\partial \tau_I \partial \tau_I} &=&  0\ , \\
\frac{1}{4}\,   \frac{\partial^2 K^{(1)|(1,l)}(\tau)}{\partial \tau_a \partial \tau_a} &=& 0\ , \\
\frac{1}{4}\,   \frac{\partial^2 K^{(1)|(1,l)}(\tau)}{\partial \tau_I \partial \tau_I} &=& 
\frac{\alpha_{II}^{(1,l)}+\alpha_{JJ}^{(1,l)} - 2 \alpha_{IJ}^{(1,l)}}{\tau_I^3 \tau_J} \ .
\ee 
The off-diagonal components are obtained from \eqref{K1_N=2}, again together with \eqref{KK-beta-alpha_aa}-\eqref{W-alpha_ab} and \eqref{N=2_Gcc_0}, and the only non-vanishing conditions are 
\be
\frac{1}{4}\,\frac{\partial^2 K^{(1)|(-1,l)}(\tau)}{\partial \tau_a \partial \tau_b} &=&  
\frac{\alpha_{aa}^{(-1,l)}+\alpha_{bb}^{(-1,l)} - 2 \alpha_{ab}^{(-1,l)}}{2 \tau_a^2 \tau_b^2} \,,\\
\frac{1}{4}\,\frac{\partial^2 K^{(1)|(1,l)}(\tau)}{\partial \tau_I \partial \tau_J} &=& 
\frac{\alpha_{II}^{(1,l)}+\alpha_{JJ}^{(1,l)} - 2 \alpha_{IJ}^{(1,l)}}{2 \tau_I^2 \tau_J^2} \,.
\ee
Obviously, all these equations can be solved by the results given in \eqref{K_N=2_KK} and \eqref{K_N=2_W}.


\section{Details concerning field redefinitions of the $\mathbb{Z}_6'$ orientifold}
\label{redef_Z6'}

Using \eqref{alpha-Y} and \eqref{Y00}-\eqref{Y33} with the constraints \eqref{tilde_G_tt_ll}-\eqref{dE/dt_i}, the $\tau$-independent (but $U_l$-dependent) coefficients 
appearing in the field redefinitions \eqref{Z6-tau-0}-\eqref{Z6-tau-3} read
\be
2\left(\alpha_{22}^{(-1,2)} - 2 \alpha_{02}^{(-1,2)} \right) = t_2 \left( - \frac{3 \delta E^{(-1,2)}}{2} +6 t_1^2 \overline{G}_{t_1 t_1}^{(1)|(-1,2)}  +t_1 \overline{G}_{t_0 t_1}^{(1)|(-1,2)}
+  t_2 \overline{G}_{t_0 t_2}^{(1)|(-1,2)} 
+ t_3 \overline{G}_{t_0 t_3}^{(1)|(-1,2)}   \right) \,, \nonumber \\
\ee
\be
2\left(\alpha_{33}^{(-1,3)} - 2 \alpha_{03}^{(-1,3)} \right) = t_3 \left( - \frac{3 \delta E^{(-1,3)}}{2} +6 t_1^2 \overline{G}_{t_1 t_1}^{(1)|(-1,3)}
+t_1 \overline{G}_{t_0 t_1}^{(1)|(-1,3)}
+  t_2 \overline{G}_{t_0 t_2}^{(1)|(-1,3)} 
+ t_3 \overline{G}_{t_0 t_3}^{(1)|(-1,3)}     \right) \,, \nonumber \\
\ee
\be
4 \alpha_{03}^{(1,2)} = t_2^{-1} \left( - \frac{\delta E^{(1,2)}}{2} +\frac{\overline{G}_{t_0 t_0}^{(1)|(1,2)}}{2} - t_1 \overline{G}_{t_0 t_1}^{(1)|(1,2)}
-  t_2 \overline{G}_{t_0 t_2}^{(1)|(1,2)} 
- t_3 \overline{G}_{t_0 t_3}^{(1)|(1,2)}\right) \,,
\ee
\be
4 \alpha_{01}^{(-1,2)} = t_2 \left( - \frac{\delta E^{(-1,2)}}{2} +\frac{\overline{G}_{t_0 t_0}^{(1)|(-1,2)}}{2} - t_1 \overline{G}_{t_0 t_1}^{(1)|(-1,2)}
+ t_2 \overline{G}_{t_0 t_2}^{(1)|(-1,2)} 
+ t_3 \overline{G}_{t_0 t_3}^{(1)|(-1,2)}\right) \,,
\ee 
\be
4 \alpha_{01}^{(-1,3)} = t_3 \left( - \frac{\delta E^{(-1,3)}}{2} +\frac{\overline{G}_{t_0 t_0}^{(1)|(-1,3)}}{2} - t_1 \overline{G}_{t_0 t_1}^{(1)|(-1,3)}
+ t_2 \overline{G}_{t_0 t_2}^{(1)|(-1,3)} 
+ t_3 \overline{G}_{t_0 t_3}^{(1)|(-1,3)}\right) \,,
\ee 
\be
2\left(\alpha_{33}^{(1,2)} - 2 \alpha_{13}^{(1,2)} \right) = t_2^{-1} \left( - \frac{3 \delta E^{(1,2)}}{2} +6 t_1^2 \overline{G}_{t_1 t_1}^{(1)|(1,2)}  +t_1 \overline{G}_{t_0 t_1}^{(1)|(1,2)}
-  t_2 \overline{G}_{t_0 t_2}^{(1)|(1,2)} 
- t_3 \overline{G}_{t_0 t_3}^{(1)|(1,2)}   \right) \,, \nonumber \\
\ee
\be
2\left(\alpha_{00}^{(-1,2)} - 2 \alpha_{02}^{(-1,2)} \right) = t_2 \left( - \frac{3 \delta E^{(-1,2)}}{2} +6 t_1^2 \overline{G}_{t_1 t_1}^{(1)|(-1,2)}  - t_1 \overline{G}_{t_0 t_1}^{(1)|(-1,2)}
+  t_2 \overline{G}_{t_0 t_2}^{(1)|(-1,2)} 
- t_3 \overline{G}_{t_0 t_3}^{(1)|(-1,2)}   \right) \,, \nonumber \\
\ee
\be
4 \alpha_{02}^{(-1,3)} = t_3 \left( - \frac{\delta E^{(-1,3)}}{2} +\frac{\overline{G}_{t_0 t_0}^{(1)|(-1,3)}}{2} + t_1 \overline{G}_{t_0 t_1}^{(1)|(-1,3)}
- t_2 \overline{G}_{t_0 t_2}^{(1)|(-1,3)} 
+ t_3 \overline{G}_{t_0 t_3}^{(1)|(-1,3)}\right) \,,
\ee 
\be
4 \alpha_{23}^{(1,2)} = t_2^{-1} \left( - \frac{\delta E^{(1,2)}}{2} +\frac{\overline{G}_{t_0 t_0}^{(1)|(1,2)}}{2} + t_1 \overline{G}_{t_0 t_1}^{(1)|(1,2)}
-  t_2 \overline{G}_{t_0 t_2}^{(1)|(1,2)} 
+ t_3 \overline{G}_{t_0 t_3}^{(1)|(1,2)}\right) \,,
\ee
\be
4 \alpha_{03}^{(-1,2)} = t_2 \left( - \frac{\delta E^{(-1,2)}}{2} +\frac{\overline{G}_{t_0 t_0}^{(1)|(-1,2)}}{2} + t_1 \overline{G}_{t_0 t_1}^{(1)|(-1,2)}
+ t_2 \overline{G}_{t_0 t_2}^{(1)|(-1,2)} 
- t_3 \overline{G}_{t_0 t_3}^{(1)|(-1,2)}\right) \,,
\ee 
\be
2\left(\alpha_{00}^{(-1,3)} - 2 \alpha_{03}^{(-1,3)} \right) = t_3 \left( - \frac{3 \delta E^{(-1,3)}}{2} +6 t_1^2 \overline{G}_{t_1 t_1}^{(1)|(-1,3)}
- t_1 \overline{G}_{t_0 t_1}^{(1)|(-1,3)}
-  t_2 \overline{G}_{t_0 t_2}^{(1)|(-1,3)} 
+ t_3 \overline{G}_{t_0 t_3}^{(1)|(-1,3)}     \right) \,, \nonumber \\
\ee
\be
2\left(\alpha_{11}^{(1,2)} - 2 \alpha_{13}^{(1,2)} \right) = t_2^{-1} \left( - \frac{3 \delta E^{(1,2)}}{2} +6 t_1^2 \overline{G}_{t_1 t_1}^{(1)|(1,2)}  - t_1 \overline{G}_{t_0 t_1}^{(1)|(1,2)}
-  t_2 \overline{G}_{t_0 t_2}^{(1)|(1,2)} 
+ t_3 \overline{G}_{t_0 t_3}^{(1)|(1,2)}   \right) \,. \nonumber \\
\ee


\section{$\mathcal{N}=2$ model: $\mathbb{T}^2 \times {\rm K}3$}
\label{sec:t2k3}

In this appendix we apply the same method as in the main text to a genuine ${\cal N}=2$ compactification, i.e.\ type I on $\mathbb{T}^2 \times {\rm K}3$. That case is more constrained via the higher supersymmetry and the results for the K\"ahler potential and for the structure of the field redefinitions are already known from the work of \cite{Antoniadis:1996vw,Berg:2005ja} (whose approaches differ from ours).  

Concretely, we consider the ${\rm K}3$ manifold at the $\mathbb{Z}_2$-orbifold point, i.e.\ ${\rm K}3 = \mathbb{T}^4 / \mathbb{Z}_2$. Obviously, there are no ${\cal N}=1$ sectors and the only ${\cal N}=2$ sector has $(m,l)=(-1,1)$ in the notation of the main text, cf.\ section \ref{ansatzN2}, meaning that it arises from closed string winding states.  

In this case, there are three relevant moduli, the 4-dimensional dilaton, called $\phi_4$ in \cite{Antoniadis:1996vw}, the volume of the $\mathbb{T}^2$, called $\sqrt{G}$ and the volume of the ${\rm K}3$, called $\omega^4$. Both of these volumes are measured with the 10-dimensional string-frame metric. In order to make contact to the notation of the main text, we rename 
\be
t_0  \leftrightarrow \phi_4, \quad t_1 \leftrightarrow \sqrt{G}, \quad  t_2 \leftrightarrow \omega^2\ .
\ee 
Out of these three fields, one can form the imaginary parts of two vector multiplet scalars, called $S_2$ and $S_2'$ in \cite{Antoniadis:1996vw}, as well as the 6-dimensional dilaton $\phi_6$, which sits in a hypermultiplet. Again, in order to make contact to the notation of the main text, we rename these fields according to
\be
\tau_0^{(0)} & \leftrightarrow & S_2 = e^{- \phi_4} G^{1/4} \omega^2 = e^{- t_0} t_1^{1/2} t_2\ , \label{tau0S2} \\
\tau_1^{(0)} & \leftrightarrow & S'_2 = e^{- \phi_4} G^{1/4} \omega^{-2} = e^{- t_0} t_1^{1/2} t_2^{-1}\ , \label{tau1S2'} \\
\tau_2^{(0)} & \leftrightarrow & e^{- 2 \phi_6} = e^{- 2 \phi_{10}} \omega^4 = e^{-2 t_0} t_1^{-1}  \label{tau2phi6}
\ee
or equivalently
\be
e^{2 t_0} &=& \frac{1}{\sqrt{\tau_0^{(0)} \tau_1^{(0)} \tau_2^{(0)}}} \ ,\label{K3_Phi-tau} \\
t_1  &=& \tau_1^{(0)}  \sqrt{\frac{\tau_0^{(0)}}{\tau_1^{(0)} \tau_2^{(0)} }} \ , \label{K3_t1-tau0} \\
t_2  &=&   \sqrt{\frac{\tau_0^{(0)}}{\tau_1^{(0)}}} \ . \label{K3_t2-tau0}
\ee   
The real parts of the complex vector multiplet scalars $S$ and $S'$ are given by the scalar dual to $C_{\mu \nu}$ and the scalar arising from the components of $C_2$ along the torus $\mathbb{T}^2$, respectively. We will denote these two scalars as $c_0$ and $c_1$. The argument of the main text for the non-redefinition of these scalars at 1-loop also holds in this genuine ${\cal N}=2$ theory.  

The form of the kinetic terms of $t_0,t_1,t_2$ and $c_0,c_1$ in the different frames is exactly as in \eqref{1-loop-EF-action-string-1}, \eqref{1-loop-EF-action-2} and \eqref{1-loop-EF-action-string_frame}, with the obvious adjustment for the range of the summations, and the metrics are again related by \eqref{Einstein_frame_Gtt} and \eqref{Einstein_frame_Gcc}. The main difference from the $\mathcal{N}=1$ case considered in the main text is the tree-level kinetic term of $t_2$ which is given by
\be
G^{(0)}_{t_2 t_2}= \frac{1}{2 t_2^2}
\ee
in contrast to the $\frac{1}{4 t_2^2}$ of equation \eqref{treelevelmetrics}. This is consistent with the kinetic term of $\omega$ in (3.1) of \cite{Antoniadis:1996vw}, obtained there by direct dimensional reduction.  


\subsection{Change of variables from $t$ to $\tau$}

The change of variables from $t_i$ to $\tau_i^{(0)}$ in the kinetic terms proceeds along the same lines as described in appendix \ref{app:ttau} for the case of an ${\cal N}=1$ orientifold. Concretely we start by introducing 
\be
& x_0= 2 t_0,\quad  x_i = \ln t_i, \qquad i \in \{1, 2 \} \,,\\
& y_i=  \ln \tau_i^{(0)} , \qquad i \in \{0, 1, 2 \}  \,.
\ee
The relation between the $x$ and $y$ coordinates is now
\be
\begin{pmatrix} x_0 \\ x_1 \\ x_2 
\end{pmatrix}= A 
\begin{pmatrix} y_0 \\ y_1 \\ y_2 
\end{pmatrix}\ ,
\ee
with $A$ being the constant  matrix given by
\be
A= \frac{1}{2} 
\begin{pmatrix}
-1 & -1 & -1 \\
 1 &  1 & -1  \\
 1 & -1 &  0
\end{pmatrix} \,. \label{K3_A}
\ee

The change of variables from $t_i$ to $\tau_i^{(0)}$ in the kinetic terms follows exactly the steps given in \eqref{changettau}. The only difference is that the sums only run until $2$ instead of $3$ and the occuring matrices are now given by
\be
X^{(0)} = \frac{1}{4}
\begin{pmatrix}
 1 & 0 &0 \\
 0& 1 & 0 \\
 0 & 0 & 2 
\end{pmatrix} \ , 
\qquad X^{(1)} =
\begin{pmatrix}
\frac{  G^{(1)}_{t_0 t_0}}{4} & \frac{t_1 G^{(1)}_{t_0 t_1}}{2} &\frac{t_2 G^{(1)}_{t_0 t_2}}{2}  \\
\frac{t_1 G^{(1)}_{t_0 t_1}}{2} & t_1^2 G^{(1)}_{t_1 t_1} & t_1 t_2 G^{(1)}_{t_1 t_2}  \\
\frac{t_2 G^{(1)}_{t_0 t_2}}{2} &  t_1 t_2 G^{(1)}_{t_1 t_2} & t_2^2 G^{(1)}_{t_2 t_2}  
\end{pmatrix} \ , \label{K3_X0X1}
\ee
\be
Y^{(0)} = A^T X^{(0)} A = 
\frac{1}{4}
\begin{pmatrix}
 1 & 0 &0 \\
 0& 1 & 0 \\
 0 & 0 & 1/2 
\end{pmatrix}
\ , \qquad
Y^{(1)} = A^T X^{(1)} A \ .
\ee
The resulting moduli metrics are 
\be
G_{\tau_i^{(0)}\tau_j^{(0)}}^{(0)}(\tau^{(0)}) &=&  \frac{Y^{(0)}_{ij}}{\tau_i^{(0)} \tau_j^{(0)}} = \frac{\left(A^T X^{(0)} A\right)_{ij}}{\tau_i ^{(0)}\tau_j^{(0)}} \ ,\\
G_{\tau_i^{(0)}\tau_j^{(0)}}^{(1)}(\tau^{(0)}) &=& \frac{Y^{(1)}_{ij}}{\tau_i^{(0)} \tau_j^{(0)}} = \frac{\left(A^T X^{(1)} A\right)_{ij}}{\tau_i ^{(0)}\tau_j^{(0)}}  
\ee
and the concrete forms of $Y_{ij}^{(1)}=\left(A^T X^{(1)} A\right)_{ij}$ are
\be
Y_{00}^{(1)} &=& \frac{1}{4} \left(\frac{G_{t_0 t_0}^{(1)}}{4} - t_1 G_{t_0 t_1}^{(1)} - t_2 G_{t_0 t_2}^{(1)}  + 
t_1^2  G_{t_1 t_1}^{(1)} +t_2^2  G_{t_2 t_2}^{(1)} + 2 t_1 t_2  G_{t_1 t_2}^{(1)} \right) \ ,   \label{K3_Y00} \\
Y_{01}^{(1)} &=& \frac{1}{4} \left(\frac{G_{t_0 t_0}^{(1)}}{4} - t_1 G_{t_0 t_1}^{(1)}  + 
t_1^2  G_{t_1 t_1}^{(1)} - t_2^2  G_{t_2 t_2}^{(1)} \right) \ , \label{K3_Y01} \\
Y_{11}^{(1)} &=& \frac{1}{4} \left(\frac{G_{t_0 t_0}^{(1)}}{4} - t_1 G_{t_0 t_1}^{(1)} + t_2 G_{t_0 t_2}^{(1)}+
t_1^2  G_{t_1 t_1}^{(1)} +t_2^2  G_{t_2 t_2}^{(1)} - 2 t_1 t_2  G_{t_1 t_2}^{(1)}  \right) \ , \label{K3_Y11}  \\
Y_{02}^{(1)} &=& \frac{1}{4} \left(\frac{G_{t_0 t_0}^{(1)}}{4}  -  \frac{t_2 G_{t_0 t_2}^{(1)}}{2} -
t_1^2  G_{t_1 t_1}^{(1)}  - t_1 t_2  G_{t_1 t_2}^{(1)} \right)\ , \label{K3_Y02} \\
Y_{12}^{(1)} &=&\frac{1}{4} \left(\frac{G_{t_0 t_0}^{(1)}}{4}  +\frac{t_2 G_{t_0 t_2}^{(1)}}{2} - 
t_1^2  G_{t_1 t_1}^{(1)} + t_1 t_2  G_{t_1 t_2}^{(1)} \right) \ ,  \label{K3_Y12} \\
Y_{22}^{(1)} &=& \frac{1}{4} \left(\frac{G_{t_0 t_0}^{(1)}}{4} + t_1 G_{t_0 t_1}^{(1)}  + 
t_1^2  G_{t_1 t_1}^{(1)}   \right) \ . \label{K3_Y22}
\ee


\subsection{Field redefinitions and K\"ahler potential}

Given that $\tau_2$ is part of a hypermultiplet, it does not couple to any of the vector multiplets, in particular not to $\tau_0, \tau_1$ and $U_1$ (all $\tau$s being 1-loop corrected). We now argue that this implies that the field redefinitions of $\tau_0$ and $\tau_1$ are independent of $\tau_2$ and, thus, we can focus on the subset of fields $\tau_0$ and $\tau_1$ when discussing their field redefinitions and the resulting K\"ahler potential. The argument for this is as follows. Given that the only ${\cal N}=2$ sector has $(m,l)=(-1,1)$, according to \eqref{N=2_Gtautau_0} the correction to the $\tau^{(0)}$-metric takes the form  
\be \label{GtauitaujN=2}
G_{\tau_i^{(0)} \tau_j^{(0)}}^{(1)}(\tau) &=& \frac{\alpha_{ij}(U_1)}{\tau_i \tau_j \tau_0 \tau_1} \ , \qquad i,j \in \{ 0,1,2 \} \ .
\ee
If this were non-vanishing for a combination of $j=2$ with $i=0$ and/or $i=1$, according to \eqref{appG_tau_tau} there would have to be a mixing of $\tau_2$ with $\tau_0$ and/or $\tau_1$ in the 1-loop field redefinition so that the metric of the corrected field variables respects the factorization of the moduli metric into  hypermultiplets and vector multiplets. For the sake of concreteness, let us assume that $G_{\tau_1^{(0)} \tau_2^{(0)}}^{(1)} \neq 0$ and that the redefinition of $\tau_1$ depends on $\tau_2$. According to \eqref{GtauitaujN=2}, $G_{\tau_1^{(0)} \tau_2^{(0)}}^{(1)}$ also depends on $\tau_0$ and $U_1$. Equation \eqref{appG_tau_tau} implies that also the field redefinition of $\tau_1$ has to depend on $\tau_0$ and $U_1$ in order to cancel the off-diagonal contribution $G_{\tau_1^{(0)} \tau_2^{(0)}}^{(1)}$. However, in that case one would obtain a $\tau_2$-dependence in the metric of the vector multiplets, for instance in the component 
\be
G_{\tau_1 U_1}^{(1)} = G_{\tau_1^{(0)} U_1}^{(1)} - \frac{\partial_{U_1} \tau_1^{(1)}}{4\tau_1^2} \ ,\label{newG_tau_u}
\ee
given that $G_{\tau_1^{(0)} U_1}^{(1)}$ does not depend on $\tau_2$, cf.\ appendix C in \cite{Berg:2005ja}. This, however, is not compatible with the factorization of the moduli space and, thus, we conclude that the field redefinition of $\tau_0$ and $\tau_1$ can not depend on $\tau_2$.\footnote{Two comments are in order: First, for this argument we assume that there is no $\tau_2$-dependent redefinition of $U_1$, as that might lead to a cancellation of the $\tau_2$-dependence in $G_{\tau_1 U_1}^{(1)}$. However, such a field redefinition was also not required for the ${\cal N}=1$ theories discussed in the main text. Second, we mention that there is actually an example where a tree-level vector multiplet scalar (the overall volume modulus) has to be redefined at 1-loop by a tree-level hypermultiplet scalar (the 4-dimensional dilaton), cf.\ eq.\ (2.2) in \cite{Antoniadis:2003sw}. The difference to our case at hand is that for \cite{Antoniadis:2003sw} the correction term in the redefinition of the overall volume modulus {\it only} depends on the 4-dimensional dilaton and not on any other vector multiplet scalars so that the above argument does not apply.}

Thus, we now concentrate on the subset of fields given by $\tau_0$ and $\tau_1$. Plugging \eqref{GtauitaujN=2} (for $i,j \in \{ 0,1 \} $) and 
\be \label{GcicjN=2}
G_{c_i^{(0)} c_j^{(0)}}^{(1)}(\tau) &=& \frac{\beta_{ij}(U_1)}{\tau_i \tau_j \tau_0 \tau_1}\ ,\qquad i,j \in \{ 0,1 \} 
\ee
into \eqref{cons-cond1} and \eqref{cons-cond3} leads to the relations
\be \label{alphabetaN=2}
\alpha_{00}=\beta_{00}\ , \qquad \alpha_{11}=\beta_{11}
\ee
and 
\be \label{beta01N=2}
\beta_{01} = \frac{1}{2} (\alpha_{00}+\alpha_{11}-2 \alpha_{01})\ .
\ee
The conditions \eqref{cons-cond2} are empty (since we only have two different indices). 

Now let us solve equations \eqref{eq-ij} and \eqref{eq-ii}. Using $\alpha_{ii}=\beta_{ii}$ in \eqref{eq-ii} we have
\be
\frac{\partial \tau_i^{(1)}(\tau)}{\partial \tau_i} =0
\ee
and then, using also \eqref{alphabetaN=2} and \eqref{beta01N=2}, equation \eqref{eq-ij} can be integrated to give
\be
\tau_i^{(1)}= \frac{2 (\alpha_{jj} - 2 \alpha_{ij})}{\tau_j} \quad \text{with} \quad i\neq j
\ee
or more explicitly
\be
\tau_0^{(1)}&=& \frac{2 (\alpha_{11} - 2 \alpha_{01})}{\tau_1} \ , \label{K3_tau_0} \\
\tau_1^{(1)}&=& \frac{2 (\alpha_{00} - 2 \alpha_{01})}{\tau_0} \ . \label{K3_tau_1}
\ee
Using this and \eqref{alphabetaN=2} and \eqref{beta01N=2}, equations \eqref{K1ii} and \eqref{K1ij} can be solved by
\be
K^{(1)} = \frac{2 (\alpha_{00}+\alpha_{11}-2 \alpha_{01})}{\tau_0 \tau_1} \ . \label{K3_K}
\ee
In \eqref{K3_tau_0}-\eqref{K3_K} all the $U_1$-dependent coefficients $\alpha_{ij}$ are proportional to the non-holomorphic Eisenstein series $E_2$ (given in \eqref{Eisenstein}), i.e.
\be \label{alphaE2}
\alpha_{ij} \sim E_2(U_1)\ ,
\ee
cf.\ \cite{Berg:2005ja}. The structure of the field redefinitions \eqref{K3_tau_0} and \eqref{K3_tau_1} was already discussed in \cite{Antoniadis:1996vw} and \cite{Berg:2005ja}, respectively. Note that the results \eqref{K3_tau_0}-\eqref{K3_K} agree with the contributions from the $(k=3)$-sector of the $\mathbb{Z}_6'$-orientifold to $\tau_0^{(1)}, \tau_3^{(1)}$ and $K^{(1)}$, cf.\ \eqref{Z6-tau-0}, \eqref{Z6-tau-3} and \eqref{K-Z6}. This is due to the fact that the $(k=3)$-sector of the $\mathbb{Z}_6'$-orientifold is formally identical with the $\mathbb{T}^2 \times \mathbb{T}^4/\mathbb{Z}_2$-orientifold, just that the third torus of the $\mathbb{Z}_6'$-model plays the role of the untwisted $\mathbb{T}^2$ of the $\mathbb{T}^2 \times \mathbb{T}^4/\mathbb{Z}_2$-model (i.e.\ we have to make the replacement $t_3^{\mathbb{Z}_6'} \leftrightarrow t_1^{{\rm here}}$ and the $(m,l) = (-1,1)$-sector at hand corresponds to the $(m,l) = (-1,3)$-sector of the $\mathbb{Z}_6'$-orientifold).

Finally, the correction to the K\"ahler potential can be expressed in terms of the quantities calculable via string amplitudes. Using 
\be
\alpha_{ij}= e^{-2 t_0} t_1 Y_{ij}^{(1)} \ . \label{alpha-YN=2}
\ee
(cf.\ \eqref{alpha-Y}), \eqref{K3_Y00}-\eqref{K3_Y11}, \eqref{tau0S2}-\eqref{tau1S2'} and \eqref{Einstein_frame_Gtt} we obtain for the K\"ahler potential 
\be
K^{(1)} = 2 t_2^2 G^{(1)}_{t_2 t_2} = 2 e^{2 t_0}  \left( t_2^2 \overline G^{(1)}_{t_2 t_2}-\frac{ \delta E}{2}\right) \ . \label{K3_K_explicit}
\ee

\end{appendix}


\end{document}